\documentclass[final,3p,times,twocolumn]{elsarticle}

\usepackage[T1]{fontenc}

\usepackage{amssymb}
\usepackage{mathtools}

\usepackage{hyperref}
\hypersetup{
  colorlinks   = true, 
  urlcolor     = blue, 
  linkcolor    = blue, 
  citecolor   = blue 
}
\usepackage[caption=false]{subfig}
\usepackage[export]{adjustbox}

\usepackage{siunitx}

\ifdefined\unit\else
  \ifdefined\NewCommandCopy
    \NewCommandCopy\unit\si
  \else
    \NewDocumentCommand\unit{O{}m}{\si[#1]{#2}}
  \fi
\fi

\usepackage{float}
\usepackage{longtable}
\newcolumntype{P}[1]{>{\centering\arraybackslash}p{#1}}
\newcolumntype{L}[1]{>{\raggedright\arraybackslash}p{#1}}
\usepackage{multirow}
\usepackage{boldline}

\journal{Nuclear Instruments and Methods in Physics Research B}

\begin{document}

\begin{frontmatter}



\title{Preparation and Characterization of Thin Arsenic Targets for Stacked-Target Experiments}


\author[1,2]{Andrew S. Voyles\corref{cor1}}
\ead{asvoyles@lbl.gov}
\cortext[cor1]{Corresponding Author}
\author[1]{Morgan B. Fox\corref{cor2}}
\ead{morganbfox@berkeley.edu}
\cortext[cor2]{Corresponding Author}
\author[1]{Jonathan T. Morrell}
\author[3]{Michael P. Zach}
\author[1]{Evan K. Still}
\author[1,2]{Lee A. Bernstein}
\author[4]{Wesley D. Frey}
\author[4]{Burton J. Mehciz}

\address[1]{Department of Nuclear Engineering, University of California, Berkeley, Berkeley, California 94720, USA}
\address[2]{Lawrence Berkeley National Laboratory, Berkeley, California 94720, USA}
\address[3]{Oak Ridge National Laboratory, Oak Ridge, Tennessee 37830, USA}
\address[4]{McClellan Nuclear Research Center, University of California, Davis, Sacramento, California 95652, USA}

\begin{abstract}
Thin, uniform arsenic targets suitable for high-fidelity cross section measurements in stacked-target experiments were prepared by electrodeposition of arsenic on titanium backings from aqueous solutions. Electrolytic cells were constructed and capable of arsenic deposits ranging in mass from approximately 1--29 mg (0.32--7.2\,mg/cm$^2$, 0.57--13\,\si\um). Examination of electrodeposit surface morphology by scanning electron microscopy and microanalysis was performed to investigate the uniformity of produced targets. Brief studies of plating growth dynamics and structural properties through cyclic voltammetry were also undertaken. Alternative target fabrication approaches by vapor deposition and electrodeposition from a deep eutectic solvent were additionally conducted. We further introduce a non-destructive characterization method for thin targets by neutron activation, which is independent of neutron flux shape, environmental factors, and source geometry, while correcting for any potential scatter or absorption effects.
\end{abstract}







\end{frontmatter}

\section{Introduction}
Thin, uniform, naturally monoisotopic arsenic ($^{75}$As) targets were created in support of stacked-target proton-induced isotope production data measurements. The associated measurement campaign of \citet{Fox2021:As} required arsenic targets of 10--50\,\si\um\ thickness at a 25\,mm diameter to extract high-fidelity production cross sections. Moreover, target material in a stack must balance having enough mass to garner appropriate production during irradiation versus the energy loss and straggle of a beam propagating through thicker targets. Target uniformity is also required since the incident beam through a multi-component stack underfills the target material.

Typically, in experimental cases like \citet{Fox2021:As}, targets of interest meeting these requirements can be sourced from commercial facilities. However, arsenic is resistant to many common thin target fabrication techniques, as has been intensively explored in the semiconductor industry \cite{Wang2017:EutecticTargetFab,Chandra1988}. Arsenic sublimes on heating and cannot be casted, nor can it be subjected to cold/hot rolling due to its brittle nature and propensity for significant cracking \cite{Rohm1972,Voyles2020:WANDA,Qaim1986:AsTarg}. Arsenic is also a poor conductor, making plasma sputtering, a common technique for the preparation of thin targets, a poor option, limited to $<1$\,mg/cm$^2$ thicknesses and a mixture of arsenic and arsenic oxides in the final target. Arsenic toxicity further hinders work with bulk quantities as it is very difficult to remediate contamination \cite{ToxicHandbook}. As a result, commercial sources only offer unsuitable thick targets (5--20+\,mm thickness) or non-uniform bulk ``lumps" \cite{AlfaAesar,Goodfellow}.

Instead, the few prior arsenic-based charged-particle nuclear data investigations have required individualized local target fabrication methods \cite{Rohm1972,Breunig2017,Nozaki1979,Paans1980,Alfassi1982,Waters1973,Brodovitch1976:ProtonsPEM,
Morrison1962,Johnson1958,Bowen1962}. Many of these methods detailed in the literature involve arsenic compounds or metallic powder suspensions in solutions, still generally unsuitable for stacked-target experiments. Specifically, arsenic targets of these forms suffer from beam current limitations due to poor thermal performance, geometries not conducive for arrangement or beam transport in stacks, or require dissolution for post-irradiation analysis that introduces contaminants, mass loss, and precision errors. Only a select number of the past arsenic reaction data campaigns required pure arsenic thin foils with the same desired qualities of this present work. These choice studies found fabrication success using vacuum evaporation and electroplating \cite{Bowen1962,Johnson1958,Qaim1986:AsTarg,Mushtaq1988:ProtonsAs}, though with varying setups and materials.

In turn, we used this foundation and developed fabrication techniques of vapor deposition and electrodeposition to meet our new stacked-target needs. We further performed detailed characterization of the created targets using scanning electron microscopy, particle transmission experiments, and neutron activation analysis. In total, 26 uniform arsenic targets ranging in mass from approximately 1 to 29 mg (0.32--7.2\,mg/cm$^2$, 0.57--13\,\si\um) were suitably prepared for stacked-target experiments.

\section{Experimental Arsenic Target Fabrication}
\subsection{\label{VaporDepSec}Vapor Deposition}
The vapor deposition of arsenic was performed at Oak Ridge National Laboratory (ORNL) using arsenic powder source material, purified of As$_2$O$_3$ contamination by heating above $460^\circ$C, beyond the compound's evaporation temperature \cite{AsOxideTemp}. 

For each vapor deposit, approximately 1\,g of arsenic was placed in the bottom of a 24\,mm inside diameter soda lime glass closed bottom vial of 2\,cm length. Source mass of 1\,g represents excess arsenic per desired target but was found to give more uniform deposition coverage with similar experiment times and final deposit mass.

\begin{figure}[H]
{\includegraphics[width=1.0\columnwidth]{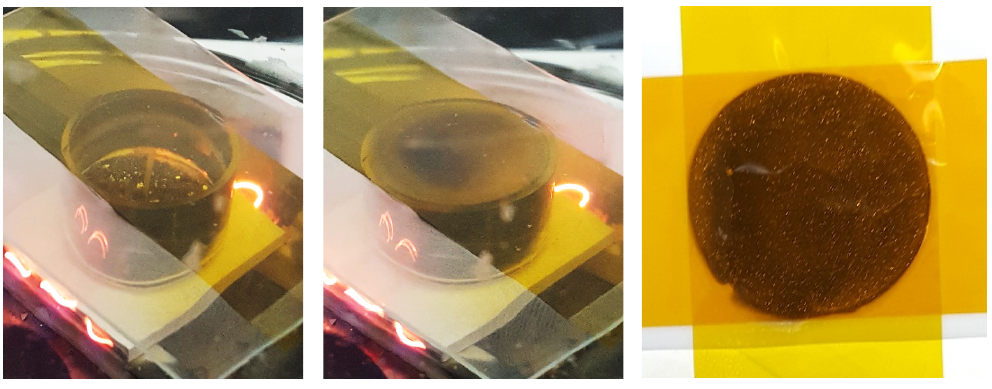}}
\vspace{-0.65cm}
\caption{Illustration of the vapor deposition process. Coil heating evaporates source arsenic within the glass vial and a thin film deposit results on the Kapton backing stuck to a microscope slide. A fully encapsulated target is the end result after cooling and removal from the vacuum chamber.}
\label{ZachProcess}
\end{figure}

The deposits in this work were prepared on 25\,\si\um\ thick Kapton film backings, which is a typical material used for sealing targets in stacked-target activations \cite{Fox2020:NbLa,Voyles2018:Nb,Marus2015,Graves2016:StackTarget}. Free-standing targets were initially explored but suffered from cracking issues during separations from the substrate. Kapton backings of at least 2.54\,cm side length were cut, massed, and fixed to a glass microscope slide (5.08$\times$7.62\,cm) by static. The microscope slide was then placed over the top of the vial opening and source mass, as pictured in Figure \ref{ZachProcess}, and held in position by gravity. The source and vial apparatus was placed on a resistive wire heating coil made from 7 passes of 24-gauge Kanthal KA1 alloy wire running through a 4\,mm thick HBN grade boron nitride insulator, capable of delivering approximately 105\,W.

\begin{figure}[H]
\centering
{\includegraphics[width=0.6\columnwidth]{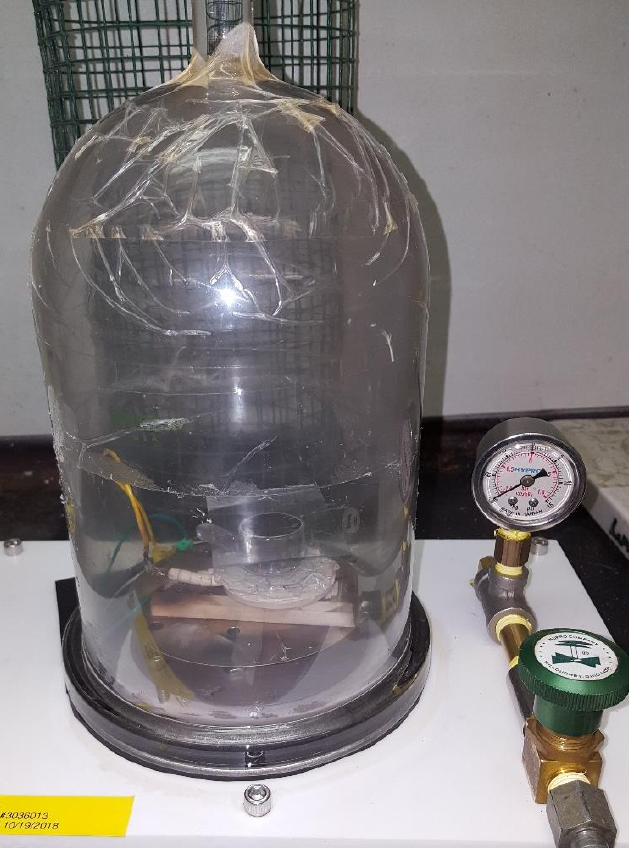}}
\vspace{-0.35cm}
\caption{Overview of vapor deposition system contained within a bell jar cover. The glass vial and microscope slide over the top of the coil heater is visible within the jar.}
\label{ZachBellJar}
\end{figure}

A 12$\times$22\,cm Pyrex bell jar cover vacuum chamber was constructed to enclose the equipment and the total deposition system is shown in Figure \ref{ZachBellJar}. Key components of the vacuum chamber were the Corian baseplate, Welch DryFast Diaphragm Pump 2034 capable of 9 torr ultimate pressure with 25 L/min free air displacement, a Nupro valve, and a 30{\textquotedbl} Hg/15 psi Bordon gauge. These were dedicated equipment for the deposition since the relatively low vapor pressure and toxicity of arsenic prohibits use in instruments that are used for other materials.

\vspace{-0.15cm}
\vspace{-0.3cm}

\vspace{-0.15cm}
\vspace{-0.3cm}

In the deposition procedure, the system was first pumped down for at least 10 minutes before the Nupro valve was closed. The heater was then operated at 15.4\,V and 7.5\,A, where arsenic deposition onto the Kapton backing took approximately 1 minute (Figure \ref{ZachProcess}). The deposited arsenic reached a yellow-colored state $\approx$45\,s into the process, likely the formation of the unstable, tetrahedral "yellow arsenic" allotrope \cite{Seidl2019}, and turned black over the subsequent 15\,s. This transformation is likely consistent with a combination of deposition of the glassy "black arsenic" allotrope, along with the rapid decomposition of the unstable yellow arsenic to the "gray arsenic" allotrope, which is the most stable of the three. It was necessary to terminate heating immediately once the deposition was completely black in color else the deposited layer reached a point of induced cracking and flaking, possibly due to the annealing of the amorphous black arsenic into the gray allotrope. A cooling period of 15 minutes was necessary to prevent As$_2$O$_3$ formation before the bell jar was vented to atmosphere.

The arsenic layer and Kapton could then be removed from the glass slide, weighed, and further sealed between two pieces of Kapton tape (44\,\si\um\ silicone adhesive on 42\,\si\um\ Kapton film) to prevent any deposition movement or escape.

\begin{figure}[H]
{\includegraphics[clip,trim=9.3cm 0.75cm 3.8cm 8.23cm, width=1.0\columnwidth]{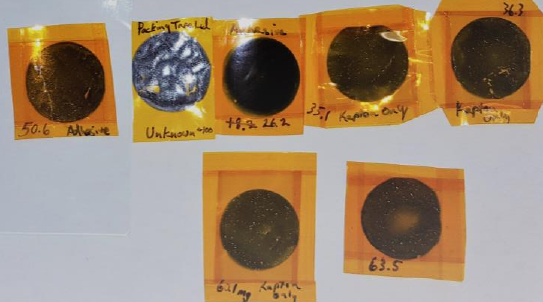}}
\vspace{-0.65cm}
\caption{Subset of final arsenic targets prepared by vapor deposition.}
\label{ZachManyTargets}
\end{figure}

This repeatable fabrication process produced 18 targets, which could be easily mounted to frames for potential use in stacked-target work. The produced targets had masses ranging from 14.4 to 119.5\,mg ($\approx$5--50\,\si\um) and a subset of them can be viewed in Figure \ref{ZachManyTargets}.

\vspace{-0.15cm}
\vspace{-0.3cm}

While vapor deposition proved to be a quick preparation method, deposition uniformity was not adequately achieved. The arsenic targets exhibited visible pinholes, thin spots, stress cracking in the more massive cases, and shape irregularities that would hinder proper nuclear data measurements. Explicitly, uniformity assessments made by transmission measurements of $^{133}$Ba and $^{88}$Y x-rays using a 2\,mm diamater pinhole collimator, at 16 different positions per target, showed variations of up to 90\% in the areal density across any given target. Some of this quantified variation is macroscopically visible in Figure \ref{ZachManyTargets}.

Replacing the Kapton backing with thin copper, titanium, or quartz may be a pathway to improved stability from vapor deposition, particularly if thicker targets (greater than approximately 10 mg/cm$^2$) are needed, however, an alternative fabrication approach for the \citet{Fox2021:As} measurement campaign was required.

\subsection{Two Electrode Electrodeposition}
\citet{Fassbender2009:TargFab} provides a brief literature review of stable arsenic electroplating methodologies and presents the experimental approach in detail for effective fabrication by electrodeposition from aqueous solutions, building from the early work of \citet{Menzies1966}. Specifically, \citet{Fassbender2009:TargFab} plates arsenic on a titanium metal backing from a solution of As$_2$O$_3$ dissolved in aqueous HCl.

The \citet{Fassbender2009:TargFab} methodology was adopted for this work but required modification to move from their 3\,mm plating radii (0.28\,cm$^2$ plating area) to targets nearly four times larger for the stacked-target application. 

\begin{figure}[H]
\centering
{\includegraphics[width=0.6\columnwidth]{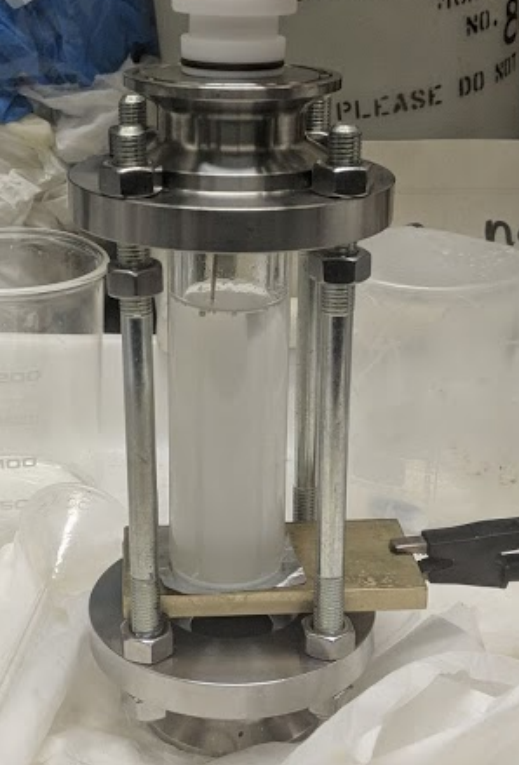}}
\vspace{-0.35cm}
\caption{Two electrode plating cell, using a titanium backing placed on a brass cathode and under a Teflon washer and the glass vial. The platinum rod anode is visible at the top of the vial, touching the surface of the plating solution.}
\label{CellInitial}
\end{figure}

Consequently, we performed electroplating at Lawrence Berkeley National Laboratory (LBNL) initially using the electrolytic cell pictured in Figure \ref{CellInitial}, powered by a Rigol DP832A DC power supply (max 30\,V, 3\,A, max power output 195\,W). The basic cell components included a platinum rod anode (diameter 1\,mm, length 10\,cm), a brass cathode support block, and a glass tube (inner diameter 23\,mm, length 10\,cm) to hold the plating solution ($\approx$41.5\,mL total available volume). All components were contained by a stainless steel base and support meant to hold the assembly in place and create a watertight seal with a nut-and-bolt compression flange.

\vspace{-0.15cm}
\vspace{-0.3cm}

The incorporated plating solution was prepared from As$_2$O$_3$ (12.5 g/L) dissolved in aqueous HCl (6M). Total solution volumes of 100\,mL were prepared by stirring the materials at $\approx$50$^\circ$C until clarified. Solution aliquots of 20--35\,mL were typically used per electroplating experiment.

Prior to the cell assembly, Ti foils of 10 and 25\,\si\um\ thickness were cut to act as backings for the arsenic electrodeposits. For each plating experiment, a Ti foil was carefully cleaned with HCl (6M) and acetone, dried and weighed, and placed on the brass cathode block beneath a Teflon washer and the glass tube. Titanium was an appropriate choice of backing in this application because it is beneficially a proton beam monitor for proton stacked-target work, presents no decay gamma-ray spectroscopy interference from activation with any eventual arsenic residual products, and is insoluble in HCl.

The cell assembly was then filled with plating solution until the platinum anode was immersed less than 1\,cm into the solution. The anode was further kept electrically isolated from the plating cell by a Teflon thermocouple adapter (inner diameter 1\,cm), which also acted to allow gases produced during electroplating (H$_2$, AsH$_3$) to be vented from the cell and avoid any potential pressure buildup.

Optimal plating times varied from 3--7 hours at a constant current supply of 130\,mA. The current value was adopted from the optimal macroscopic current density 31.2\,mA/cm$^2$ used in \citet{Fassbender2009:TargFab}. Following shutoff of the power supply, the plated targets were removed from the cell, washed with HCl (6M), dried in air, and weighed.

\subsection{Deep Eutectic Solvent Electrodeposition}
Many of the targets plated using the adopted \citet{Fassbender2009:TargFab} methodology exhibited visible flaking of arsenic dendrites, particularly for thicker targets formed by longer plating time, prompting the exploration of an alternative electrodeposition process, stemming from the work of \citet{Wang2017:EutecticTargetFab}. This alternative methodology is based upon the plating of a 0.02 M solution of As$_2$O$_3$, dissolved in a deep eutectic solvent, formed by the mixing of choline chloride and ethylene glycol in a 1:2 molar ratio, stirred at $\approx$50$^\circ$C for approximately 30 minutes until transparent. Using this version of plating solution, the two electrode plating cell of Figure \ref{CellInitial} was still employed, using the same 1\,mm diameter platinum rod as a counter electrode, but with either Cu or Al foils (both of which, like Ti, are proton beam monitor standards) of 15\,\si\um\ thickness as a working electrode. Potentiostatic deposition for this setup did not yield any visible arsenic deposits, nor did galvanostatic plating at the current densities described in the work of \citet{Wang2017:EutecticTargetFab}, despite significant gas evolution. By raising the current density to $\approx$24.1\,mA/cm$^2$, galvanostatic plating produced a robust, thick, and glossy black deposit when plating onto Al foils after approximately 2 hours. Plating onto Cu foils produced irregular, mottled deposits of green and red dendritic formations. However, later characterization (see Section \ref{SectionSEM}) revealed that these deposits contained only trace amounts of heterogeneously distributed arsenic, and instead appeared to be consistent with a combination of Al and aluminum oxides (for plating onto Al foils), or copper oxides (for plating onto Cu foils). Scanning electron microscope (SEM) micrographs in both cases revealed no evidence of any bulk deposits, but rather, a number of randomly dispersed grains (sub-\si\um\ in size), and a highly cracked surface throughout the plating area. This morphology is consistent with those micrographs presented by \citet{Wang2017:EutecticTargetFab}, who did not perform energy dispersive x-ray spectroscopy (EDS) characterization, but noted that the X-Ray Diffraction (XRD) spectra of their targets exhibited no XRD patterns due to arsenic. As a result of the conclusive lack of arsenic in this alternative plating approach, we instead revised our aqueous HCl plating methodology to produce targets capable of meeting the specifications for stacked-target applications.

\subsection{Three Electrode Electrodeposition}
\begin{figure}[H]
\centering
{\includegraphics[width=0.85\columnwidth]{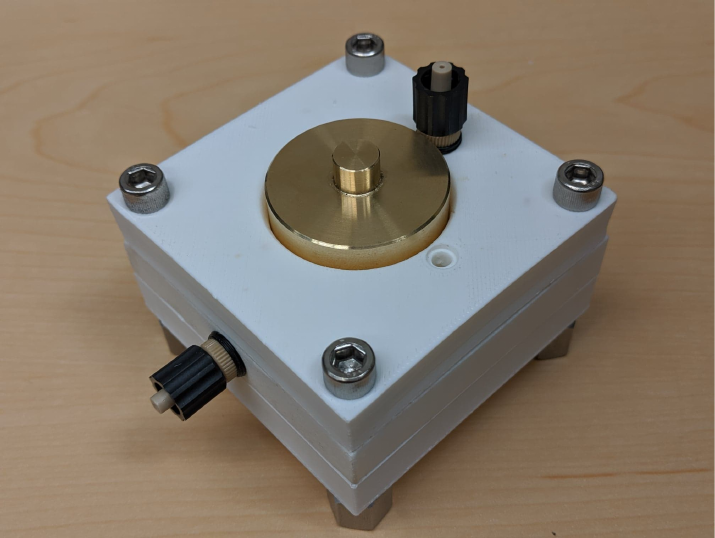}}
\vspace{-0.35cm}
\caption{Refined plating cell incorporating three electrode capability and more controlled solution handling. The brass cathode is clearly visible but the remainder of the plating components, including the titanium backing and platinum plate anode, are hidden within the cell. An opening for the Ag/AgCl reference electrode is visible to the lower right of the cathode.}
\label{RefinedCell}
\end{figure}

To this end, a refined version of the aqueous HCl electrolytic cell was additionally created in order to improve plating consistency, quality, and control. In particular, although gas venting was physically managed in the Figure \ref{CellInitial} setup, significant gas evolution was seen at the platinum rode counter electrode and it is likely that our adapted version of the \citet{Fassbender2009:TargFab} methodology was hydrolyzing a significant amount of water, quickly moving into a depletion-limited plating regime. As a result (and exacerbated by the narrow rod for the counter electrode), any free arsenic would be rapidly drawn to the working electrode, leading to the observed dendritic structures for the thicker targets (made via longer plating times). Thus, shifting to an alternative plating cell design, with a larger counter electrode placed closer to the working electrode, was expected to yield more uniform field lines, and by extension, more uniform plating for larger thicknesses. This new iteration, pictured in Figure \ref{RefinedCell}, was 3D-printed from ABS plastic. In this design, the brass support block is placed on top of the Ti cathode backing foil and an O-ring, and plating cell leg stands thread onto the base to form a watertight compression seal between the plating cell and the backing foil. The previous cell's platinum rod anode is replaced by a platinum plate and mounted opposite the O-ring in this assembly. Luer-lock connectors were also added to control the plating solution addition between the electrodes and to better maintain the cell assembly.

\vspace{-0.15cm}
\vspace{-0.3cm}

This new cell was further modified to include an Ag/AgCl reference electrode, inserted into the top of the plating cell, held in place by an O-ring. This three electrode system was powered by a Pine Instruments WaveDriver 100 potentiostat (max 24\,V, 5\,A, max power output 600\,W).

An analysis of this cell was conducted by cyclic voltammetry. The chemistry-focused explanation of results is currently underway and will be expounded in the ultimate publication. In spite of a lacking to-date detailed reflection of the chemistry basis, the voltammetry studies still provided best settings for the target fabrication. Electroplating experiments were completed using either galvanostatic mode with a current of ${<150}$\,mA, or in potentiostatic mode with a voltage of $<3$\,V, during 4--12 hours.

\begin{figure}[H]
\centering
	\subfloat[Kapton-encapsulated and frame-mounted arsenic targets ready for stacked-target experiments. The consistency among the targets is a strength of the developed electroplating process. \label{AsBNLfoils_group3}]
		{\includegraphics[width=1.0\columnwidth]{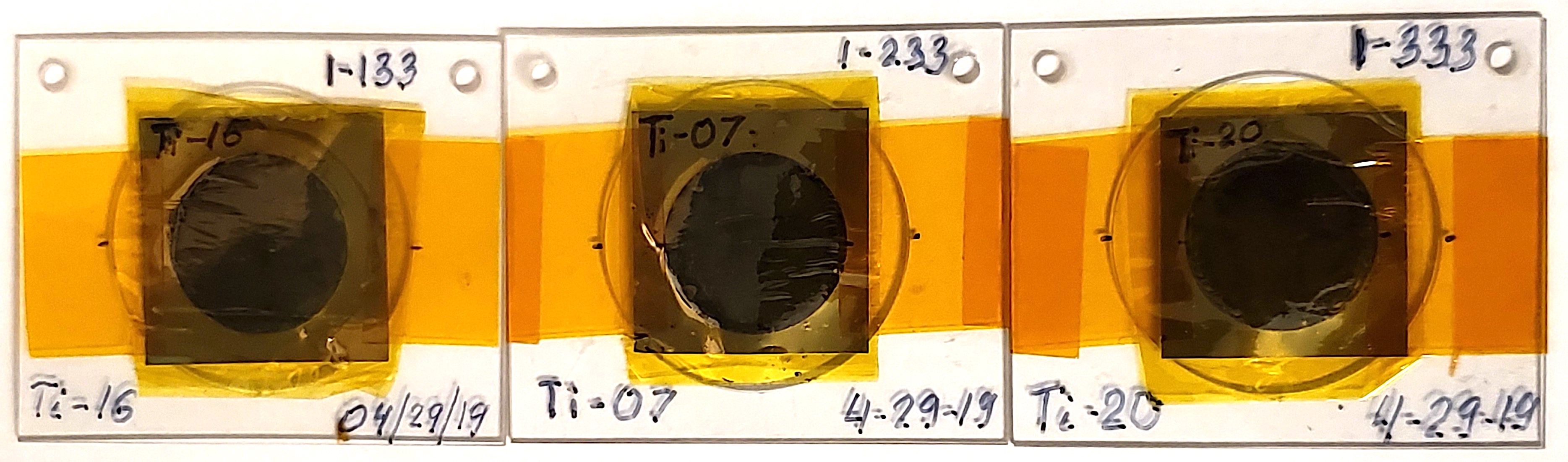}}\\
	\subfloat[Bare arsenic target prior to any encapsulation in Kapton after removal from the plating cell. The observable curved titanium backing is a typical outcome of the cell fastening and build-up of in-plane stress during the deposition process. \label{AsFoilBare3}]{\makebox[1.0\columnwidth]
		{\includegraphics[width=0.6\columnwidth]{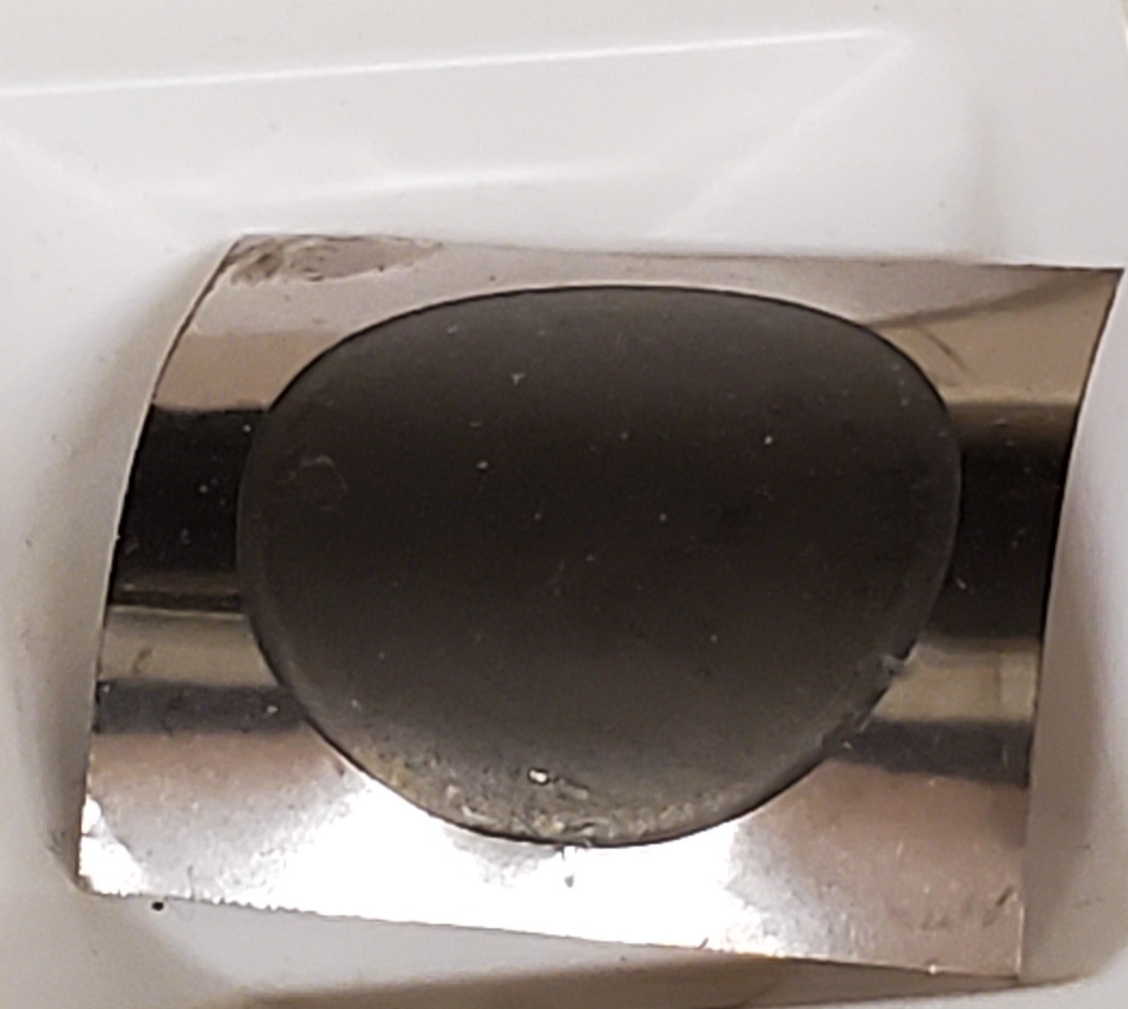}}}
	\vspace{-0.15cm}
	\caption{Representative subset of arsenic targets prepared by electrodeposition.}\label{ElectroplatedExamples}
\end{figure}

Overall, using both of the aqueous HCl electrolytic cells, we developed a consistent plating capability at LBNL and produced over fifty 22.5\,mm diameter arsenic targets through electrodeposition. The targets ranged in mass from approximately 1 to 40\,mg (0.3--10\,mg/cm$^2$), however the heavier cases developed significant stress from formed arsenic dendrites and were prone to flaking during removal and handling. Still, the overall target quality was qualitatively appropriate for stacked-target work and 26 of the most structurally-sound targets (0.32--7.2\,mg/cm$^2$) were chosen for the \citet{Fox2021:As} experiments. Representative samples from the 26 are shown in Figure \ref{ElectroplatedExamples}.

\section{\label{SectionSEM}Target Microscopy and Microanalysis}
A necessarily more detailed assessment of the plated electrodeposits' uniformity was carried out using a FEI Quanta 3D field emission gun with a focused ion beam (FIB) - SEM instrument and EDS capabilities \cite{Lupinacci2014,Frazer2015}.

\begin{figure*}[t]
\centering
	\subfloat[Imaging showing microstructure overview in center region of target. \label{SEM500um}]
		{\includegraphics[width=0.3\textwidth]{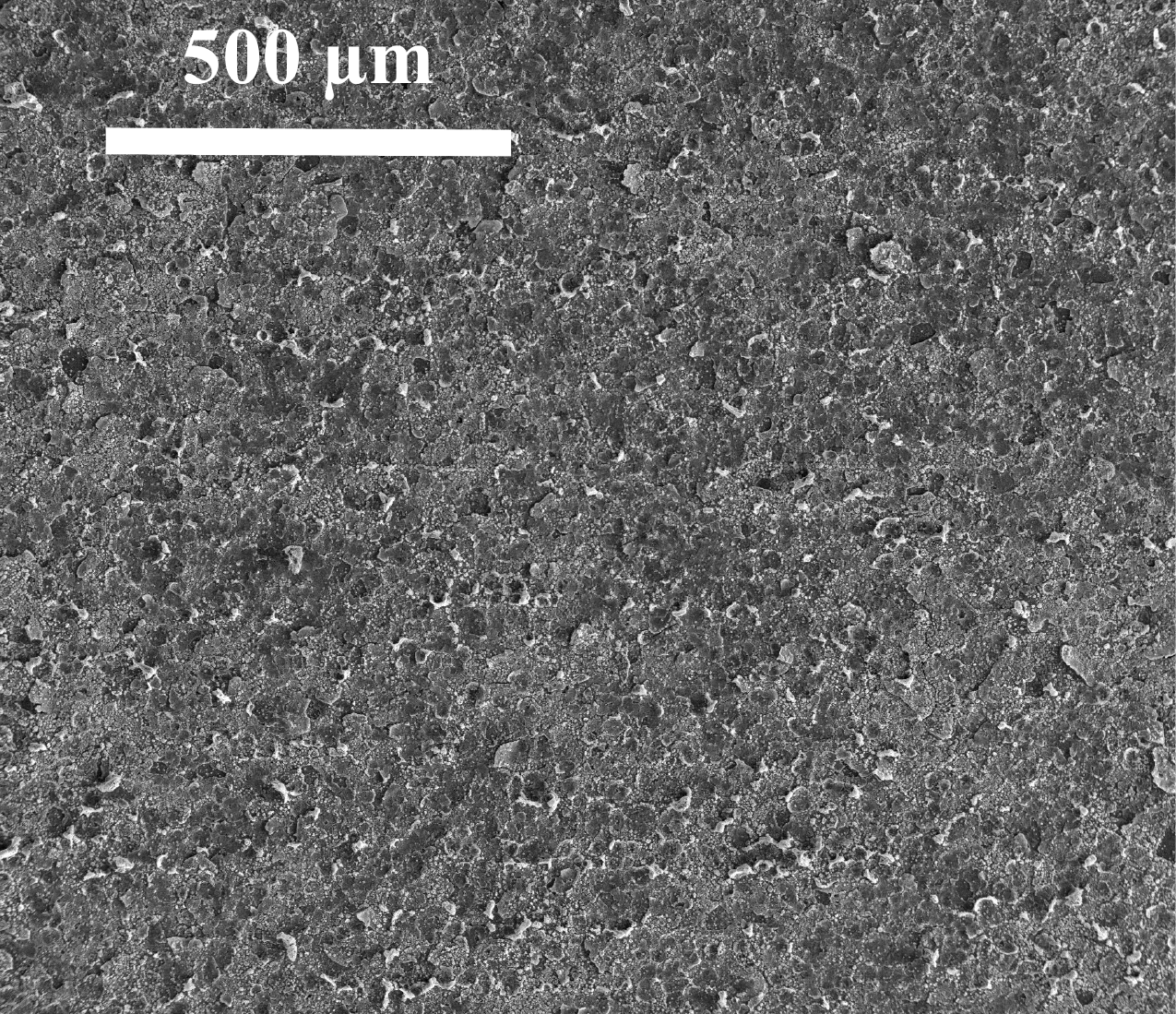}}\hspace{2mm}
	\subfloat[Increased magnification in central target area. \label{SEM300um}]
		{\includegraphics[width=0.3\textwidth]{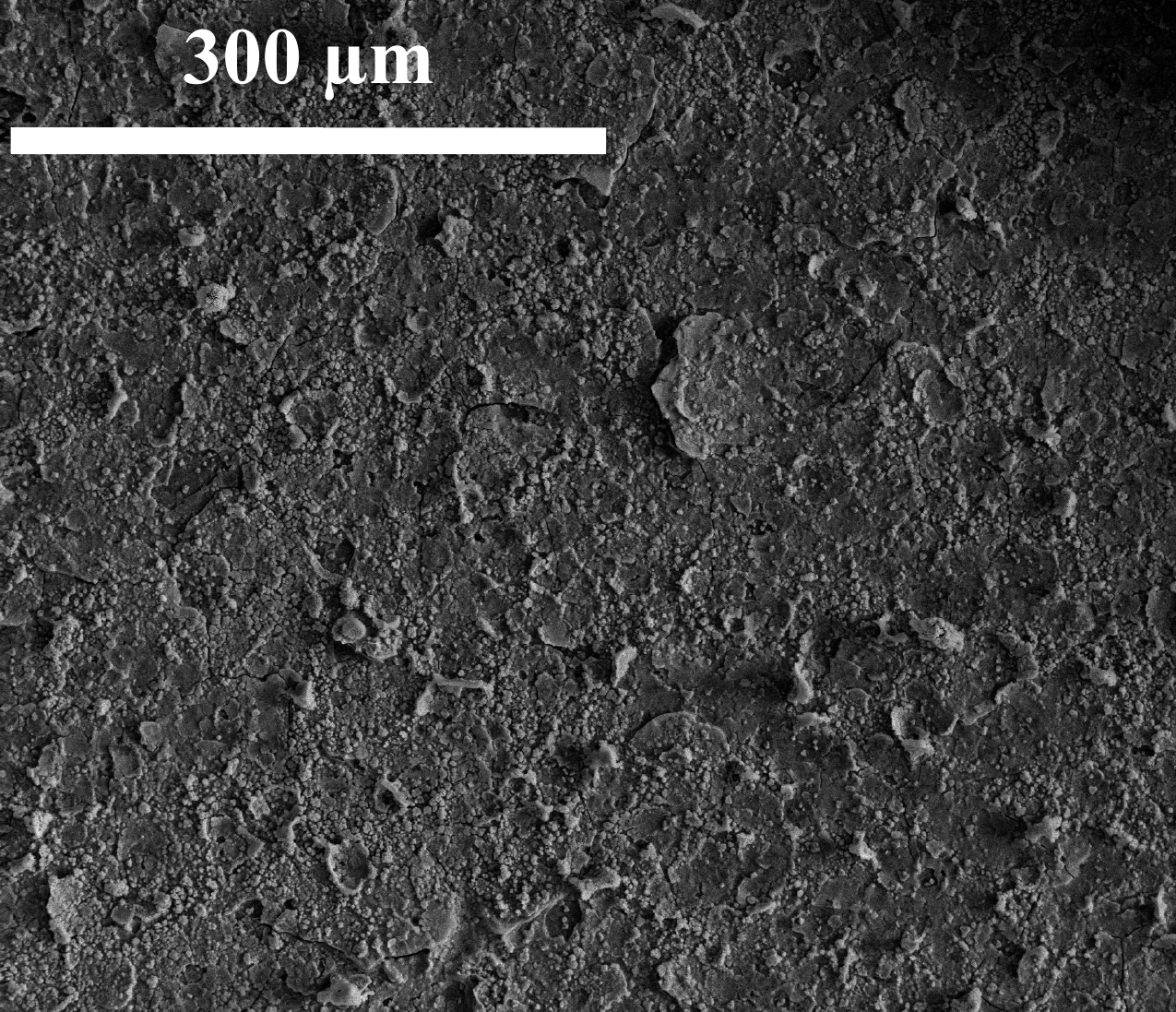}}\hspace{2mm}
	\subfloat[Maximum used magnification examining central target area. \label{SEM30um}]
		{\includegraphics[width=0.3\textwidth]{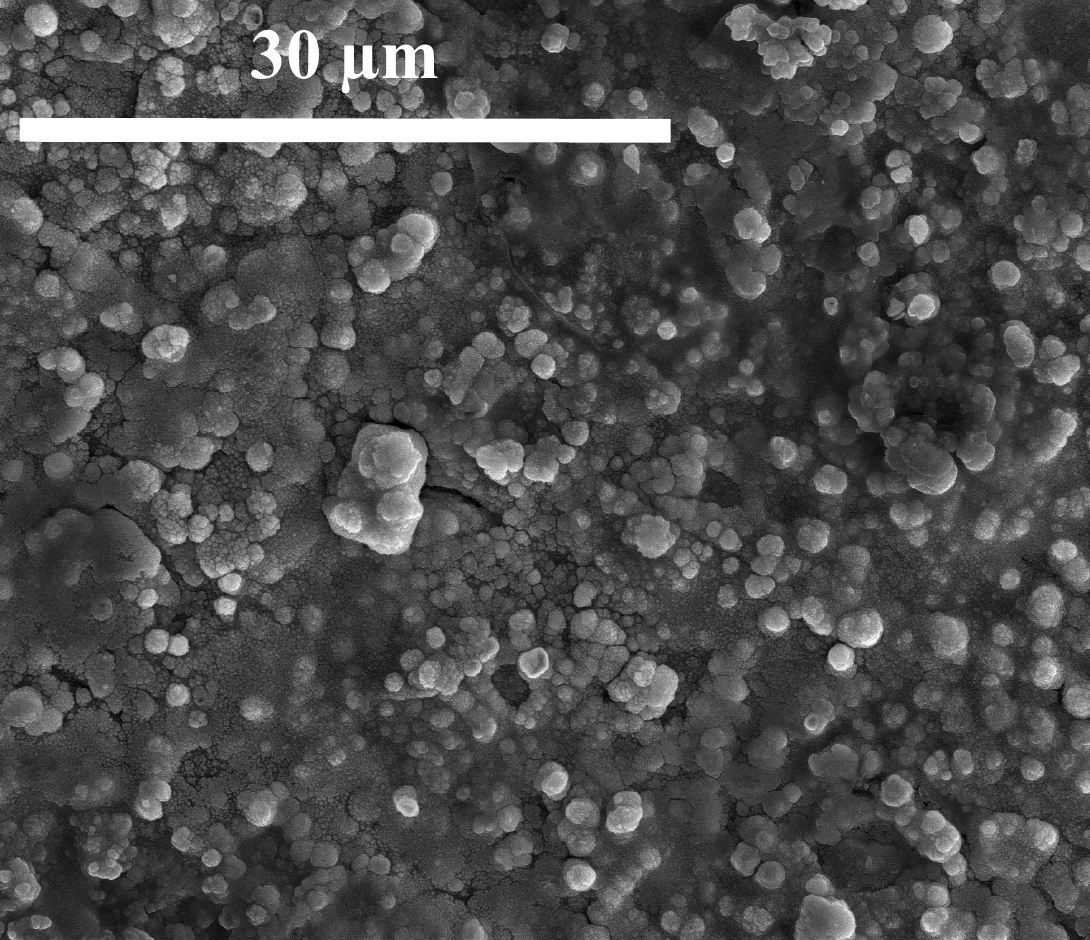}}\hfill
	\subfloat[Overview of plating edge, showing clear delineation between the arsenic layer and the titanium backing. \label{SEMedge}]
		{\includegraphics[width=0.3\textwidth]{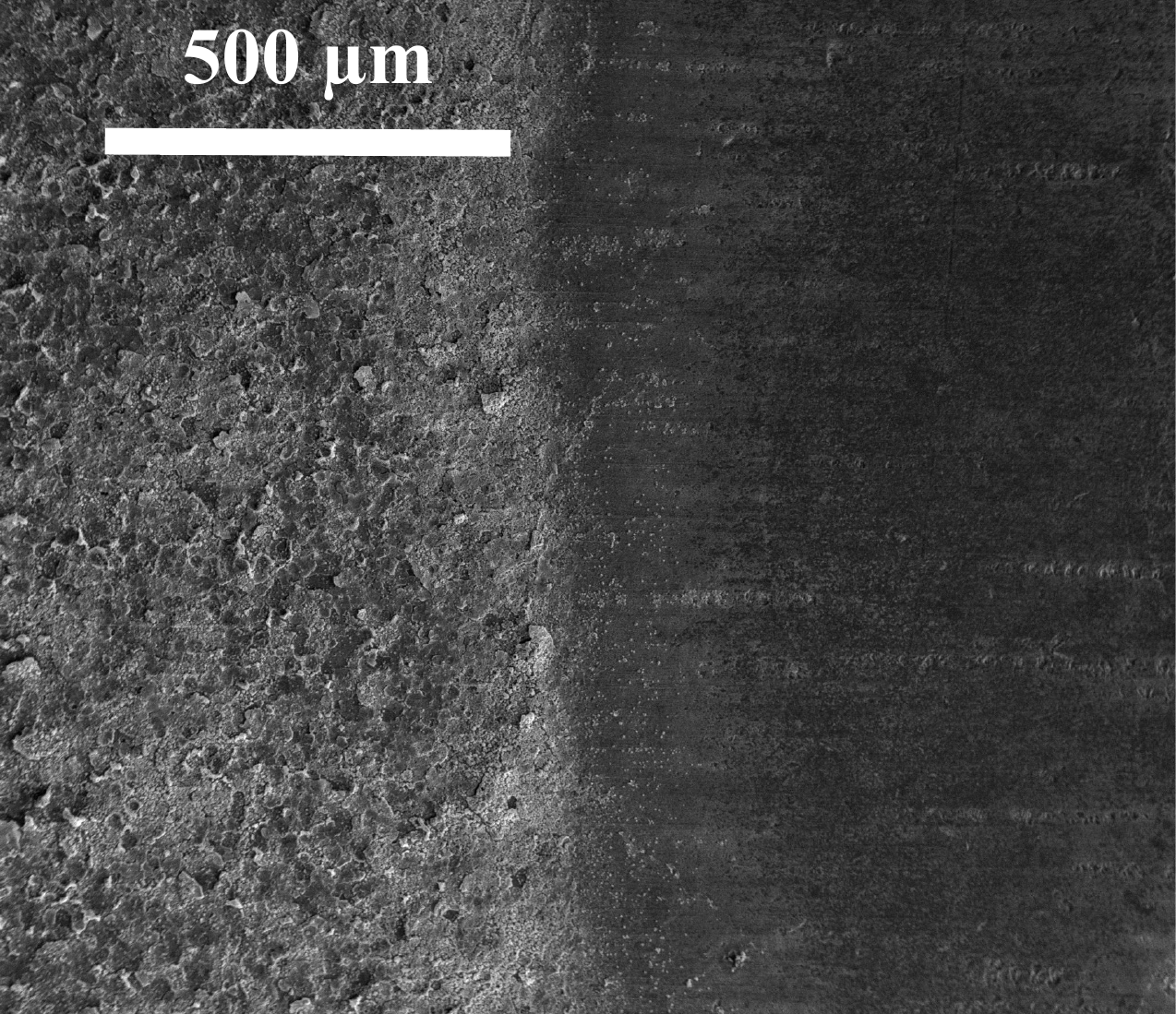}}\hspace{2mm}
	\subfloat[EDS scan along the edge of the arsenic deposition, more quantitatively imaging transition from arsenic layer to titanium backing. Some edge shape irregularity is noted but is smaller than 1\,mm. The scan further confirms the consistency and uniformity of the arsenic layer. \label{SEMedgeEDS}]
		{\includegraphics[width=0.4\textwidth]{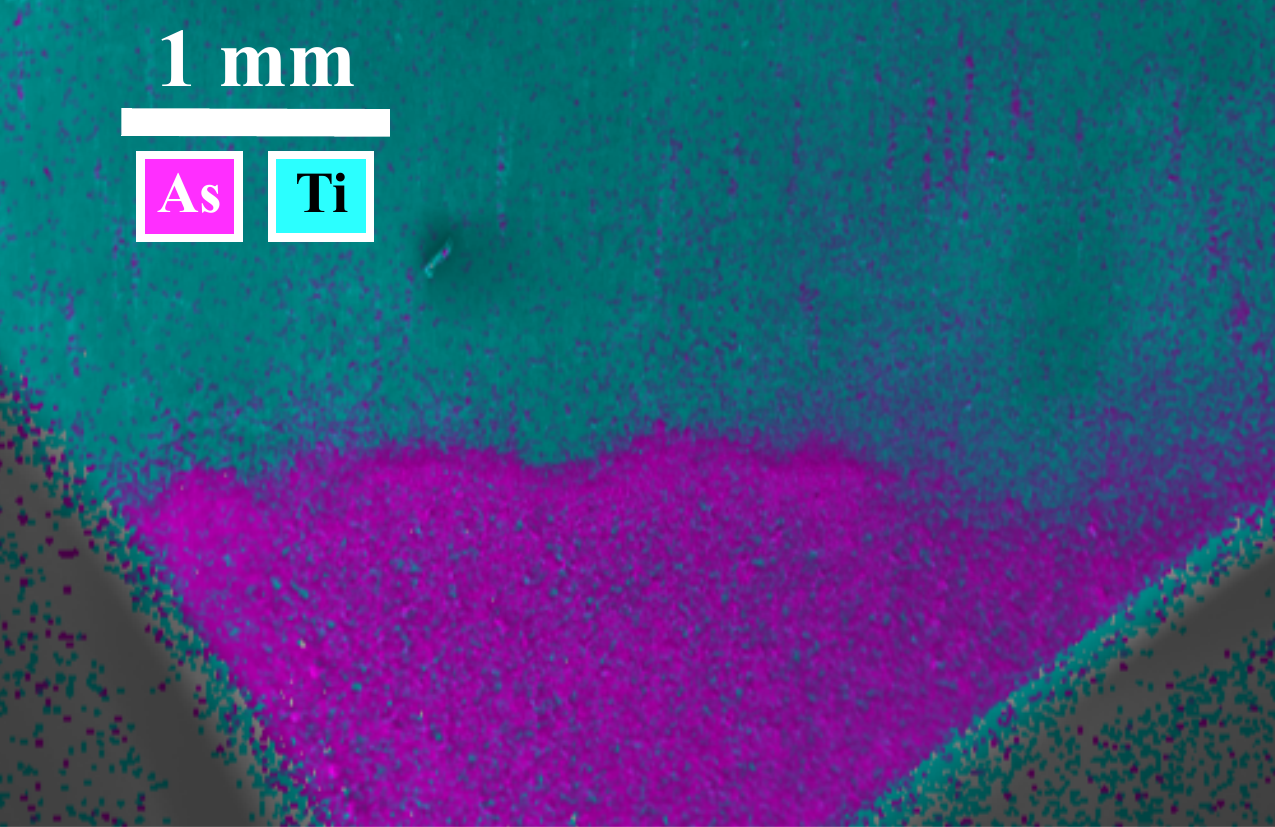}}
	\vspace{-0.15cm}
	\caption{SEM micrographs of arsenic electroplated from dissolved As$_2$O$_3$ (12.5 g/L) in aqueous HCl (6M) using plating times of 3--12 hours.}\label{As_SEM}
	\vspace{-0.35cm}
\end{figure*}

SEM imaging was performed over length scales of 1000--30\,\si\um\ for sampled targets, captured at 2\,mm increments over the surface of the arsenic depositions. Although some cracking and flaking features were present, there were clearly defined arsenic layers with overall uniform morphology at the length scales probed and no position-dependent properties save for directly at the deposition edges. Specifically, small ($<$1 mm diameter) irregularities were observed around the outside edge of targets, due to nucleation of hydrolysis gases at the interface between the backing foil and the Teflon washer. However, these minor irregularities existed far outside the expected stacked-target beam spot and were therefore not problematic. Figure \ref{As_SEM} demonstrates these SEM results.

EDS line scans and mappings at the deposition edges showed clear delineation between the Ti backing and the arsenic layer. Equivalent scans throughout the central region of the targets showed a consistent, but not perfectly smooth, layer of arsenic. A representative EDS map can be reviewed in Figure \ref{SEMedgeEDS}. Analysis of deposit thickness and the arsenic interface with Ti backings was also attempted through trenching using the Ga$^+$ ion trenching using the FIB, however, the beam effects on target prevented any clear insight. 

As a result, the microscopy and microanalysis efforts informed that the targets were uniform and consistent but the imaging was insufficient to characterize the target thickness and areal density properties to a level of precision necessary for stacked-target activation calculations. Still, the knowledge of uniformity opened the resources appropriate for further characterization work, as whole-target assessments were possible rather than focused analysis on expected beam spot positions of the electrodepositions.

\section{Target Characterization}
Although mass information was recorded prior to and following the electroplating experiments, the analytical balance (with 1\,mg resolution) proved largely insensitive to the arsenic deposition mass versus the far more massive Ti foil backings. This insensitivity resulted in variable or unphysical descriptions of the arsenic layers, which were not useful for the precision demanded from stacked-target applications. Similar issues existed for \citet{Waters1973} following their arsenic target fabrication.

Beyond the massing attempts, conventional non-destructive thickness measurements by micrometers were also unavailable here due to the safety and fragility constraints imposed by the target fabrication process. Instead, the necessary arsenic deposition areal densities had to be characterized through alternative means.

\subsection{Particle Transmission Experiments}
The most accessible substitute approach for target areal density determinations were transmission measurements. Initially, the x-rays of $^{133}$Ba, $^{88}$Y, $^{241}$Am, and $^{109}$Cd were used for the target characterization, similar to the procedure performed for the vapor deposition targets in Section \ref{VaporDepSec}. Point sources of the isotopes were combined with a narrow collimator to investigate the depositions in a grid-like manner and yield both thickness information as well as a continued check of uniformity. An ORTEC GMX series (model GMX-50220-S) High-Purity Germanium (HPGe) detector and a single leaf of an EURISYS MESURES 2-Fold Segmented Clover HPGe detector were used for this activity. 

Given that the arsenic was deposited onto titanium backings and encapsulated in Kapton tape, the transmission work on the produced electroplated targets had to be performed relative to a pure titanium and Kapton control case. The control provided a baseline transmission intensity for the different utilized x-rays. However, the significant thickness of combined titanium, kapton film, and kapton adhesive components compared to the thin deposited arsenic layers created difficulty in distinguishing the transmission through the electroplated targets versus the control. Specifically, across almost all targets, the transmitted x-ray intensities were found to be greater than or equivalent to the baseline control intensity within uncertainty since any additional arsenic attenuation provided negligible contrast. Further, the transmission work sensitivity could not be increased by using other lower energy x-rays because the relatively large areal density presented by the titanium and Kapton rendered overall photon transmission to zero. Consequently, photon transmission characterization results could not be fully formed.

In turn, charged particle transmission measurements were instead explored because the increase in stopping power versus photons was expected to lead to more pronounced intensity differences through the thin arsenic layers. Therefore, both $^{90}$Sr and $^{204}$Tl beta-emitting point sources were used with a Geiger-Muller (GM) tube and a Spectech ST360 counter to characterize the produced arsenic foils. Count data was collected with the GM tube over 10-20 minute periods for each arsenic foil at a supplied bias of 800\,V. The beta source was placed 4\,cm from the detector window of the GM tube, while a collimator sat at a 2\,cm distance from the detector window and the targets were held at a 1\,cm distance from the window. In order to determine an absolute value for the electroplated arsenic areal densities from the count data, additional transmission measurements of control materials and background were collected to generate a beta-transmission calibration curve.

The measured target areal densities by electron transmission were all physical results and did not suffer from the insensitivity present in the x-ray work. In spite of this sensitivity improvement, uncertainties for the determined areal densities existed at the 20--30\% level and there was unexpected variation in the relative magnitudes between electroplated targets made from the same experimental batch. Likely, an unaccounted-for convolution of electron stopping power through multiple materials and the sources' beta-emission spectra was degrading the results. In attempts to remediate this issue, calculations incorporating the continuous slowing down approximation range for electron attenuation, a semi-empirical transmission-based areal density formula derived from the Fermi function of beta-spectra, and Bethe-Bloch focused simulations were applied. Largely, none of these corrective efforts fixed the magnitude variation for targets or provided greater precision than the base experimental work. 

Although the theoretical basis for transmission characterization is sound, its application in practice to the prepared thin targets was not trivial. These traditional electromagnetic probes of areal density suffer here because of the significantly greater stopping in the relatively much more massive backings versus the arsenic layers. It is possible that a benefit could be gained from a more precise experimental setup or transitioning to $\alpha$-transmission work but expected improvements are unknown and may still not qualify the findings for eventual stacked-target calculations \cite{Sugai1997}.

\subsection{Neutron Activation}
Ultimately, with transmission experiments providing inadequate accuracy and precision needed for high-fidelity production cross section measurements, another alternative approach to characterizing these foils was conducted using neutron activation of the arsenic relative to known neutron capture reactions. The relative measurement allows for the arsenic target masses and uncertainties to be calculated on the basis of well-described capture cross sections. Moreover, given that the target foils were fabricated for proton beam irradiations, the capture products resulting from a neutron irradiation pose no contamination risk to any eventual proton-induced products. Still, note that for the particular arsenic targets of \citet{Fox2021:As}, the charged-particle irradiations were performed 1--2 years prior to any of the following described neutron activation work and thus it is certain that there is no contamination from capture products to the \citet{Fox2021:As} results.

To this end, neutron irradiation experiments were carried out at the UC Davis McClellan Nuclear Research Center. The McClellan site has a 2 MW TRIGA reactor with a unique capability to perform external beam neutron radiography for items such as plane wings and fuel injectors over a large area \cite{McClellanNRC}. Their smallest external imaging bay was consequently an appropriate and easily accessible setting to non-destructively characterize the arsenic targets.

\subsubsection{Experimental Setup and Procedure}
The nuclear data associated with the arsenic neutron capture reaction were critical to the development of this characterization technique. More specifically, as seen in the $^{75}$As(n,$\gamma$)$^{76}$As data shown in Table \ref{ActivationReactions}, the capture product has an appropriate half-life for gamma spectroscopy, a large enough and well-characterized thermal capture cross section to achieve sufficient activation with a small amount of mass, and an intense decay gamma-ray in a useful energy range for HPGe detection. Furthermore, the reference neutron monitor material chosen in this work had to share similar properties and were thus chosen as a function of their half-life, thermal capture cross section magnitude and uncertainty, as well as their capture products’ primary decay gamma-ray energy and intensity. Eight total neutron monitor foils were used in this investigation: 5 $^{197}$Au, 2 $^{\textnormal{nat}}$Cu, and 1 $^{\textnormal{nat}}$Fe. The relevant nuclear data associated with these monitors is also provided in Table \ref{ActivationReactions}.

\begin{figure*}[!h]
\centering
	\subfloat[Overview of shutter and target geometry in McClellan imaging bay. \label{ShutterAndFrame}]
		{\includegraphics[width=0.65\textwidth]{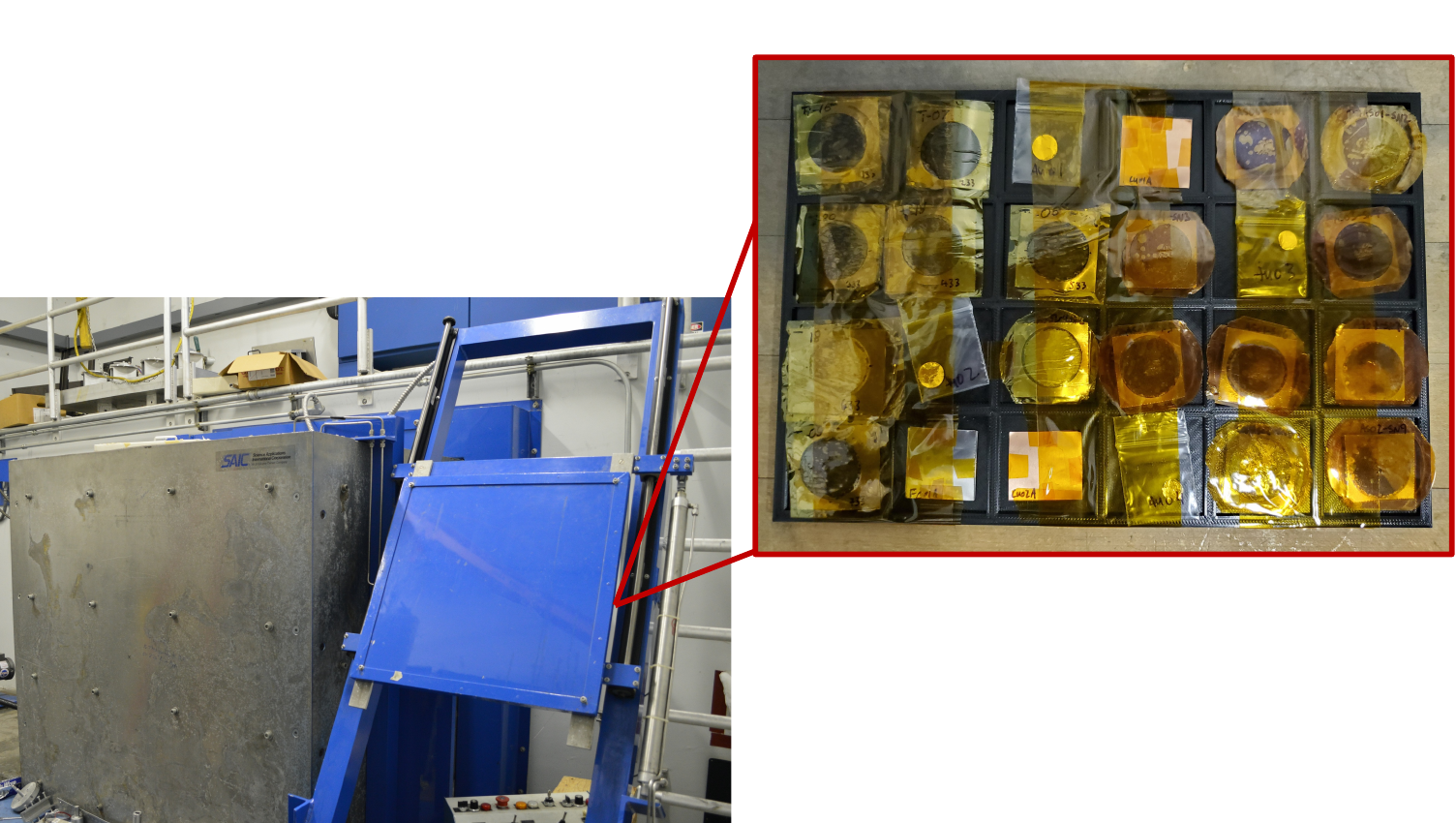}}\hspace{5mm}
	\subfloat[Direct view into beamline collimator of the imaging bay. The darkened center of the beamline is the reactor tank wall. \label{BeamlineCollimator}]
		{\includegraphics[width=0.3\textwidth]{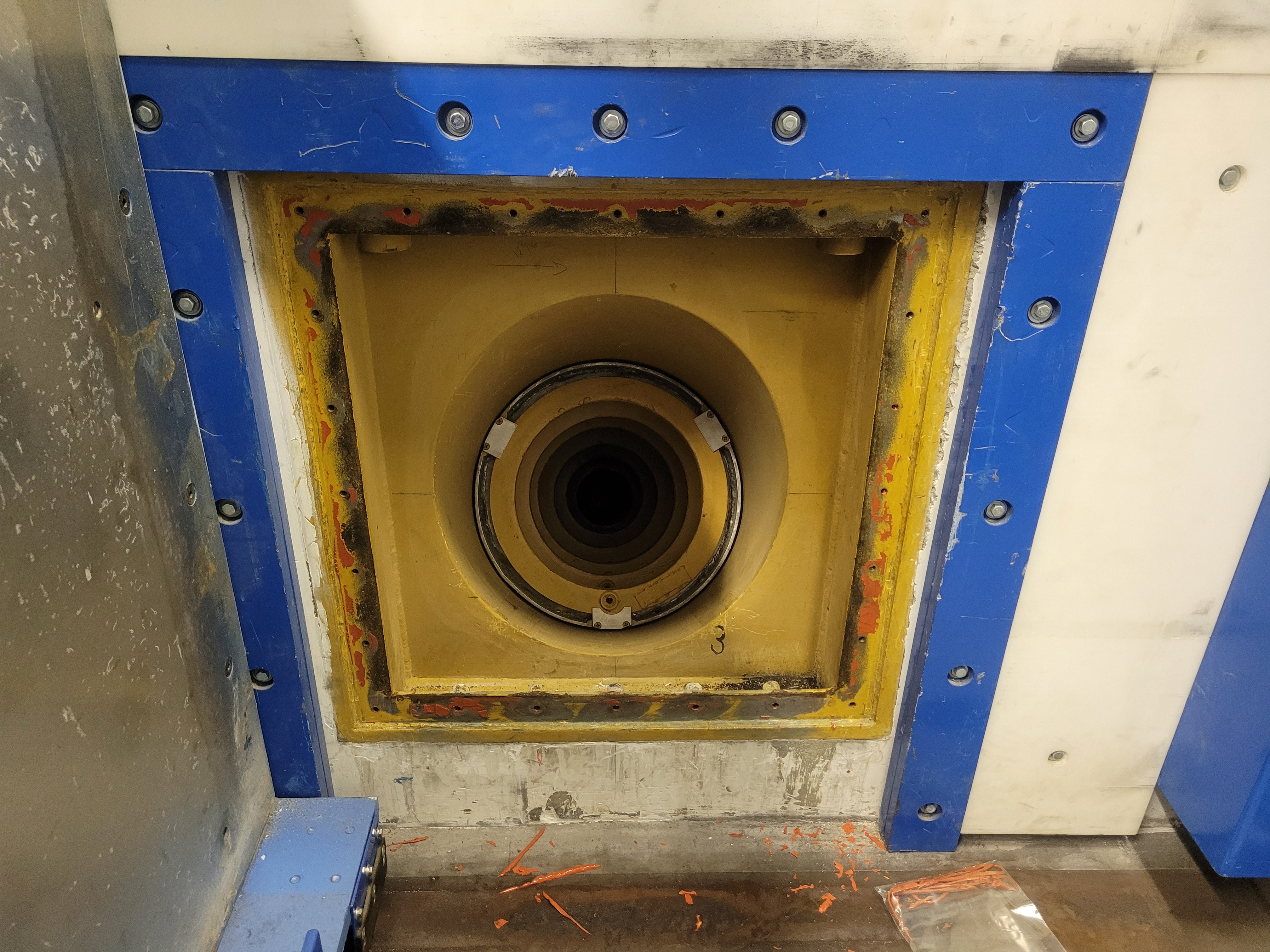}}
\caption{Experimental setup in an external imaging bay at the McClellan Nuclear Research Center. The (a) blue fast shutter, situated beside the large beam stop, moves into the neutron beamline as the beam stop moves away. The angled position of the shutter is needed to intersect the incoming (b) $30^\circ$ angled neutron beam path perpendicularly. The frame holding the arsenic targets and monitor foils is attached to the side of the fast shutter that faces the beam.}
\label{McClellanBaySetup}
\end{figure*}

\begin{figure*}
\centering
	\subfloat[Preliminary radiograph of neutron beam spot ($\approx$15-inch diameter) at eventual target frame position on fast shutter. \label{RadiographCrop}]
		{\includegraphics[width=0.3\textwidth, valign=c]{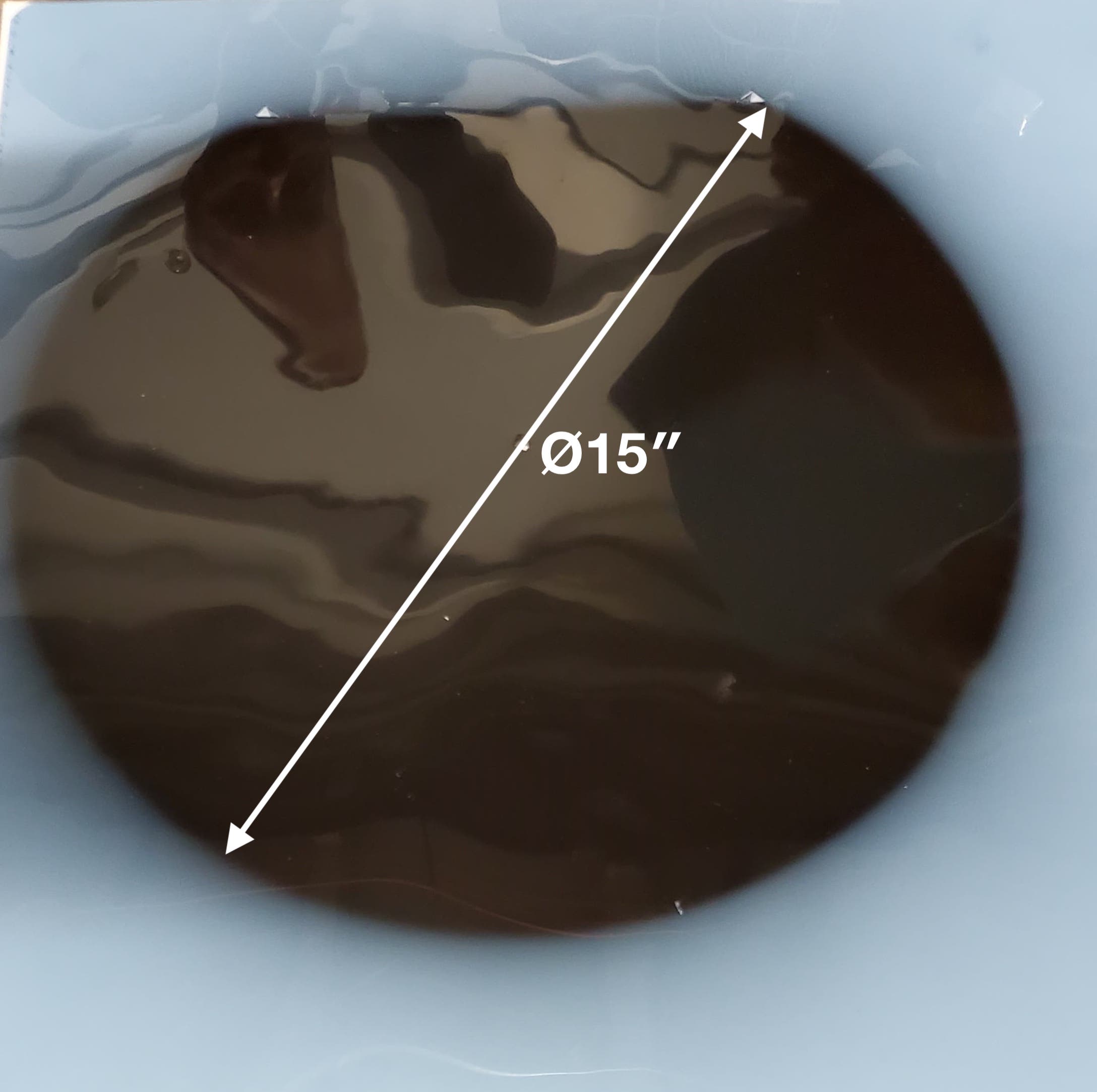}
		\vphantom{\includegraphics[width=0.65\textwidth, valign=c]{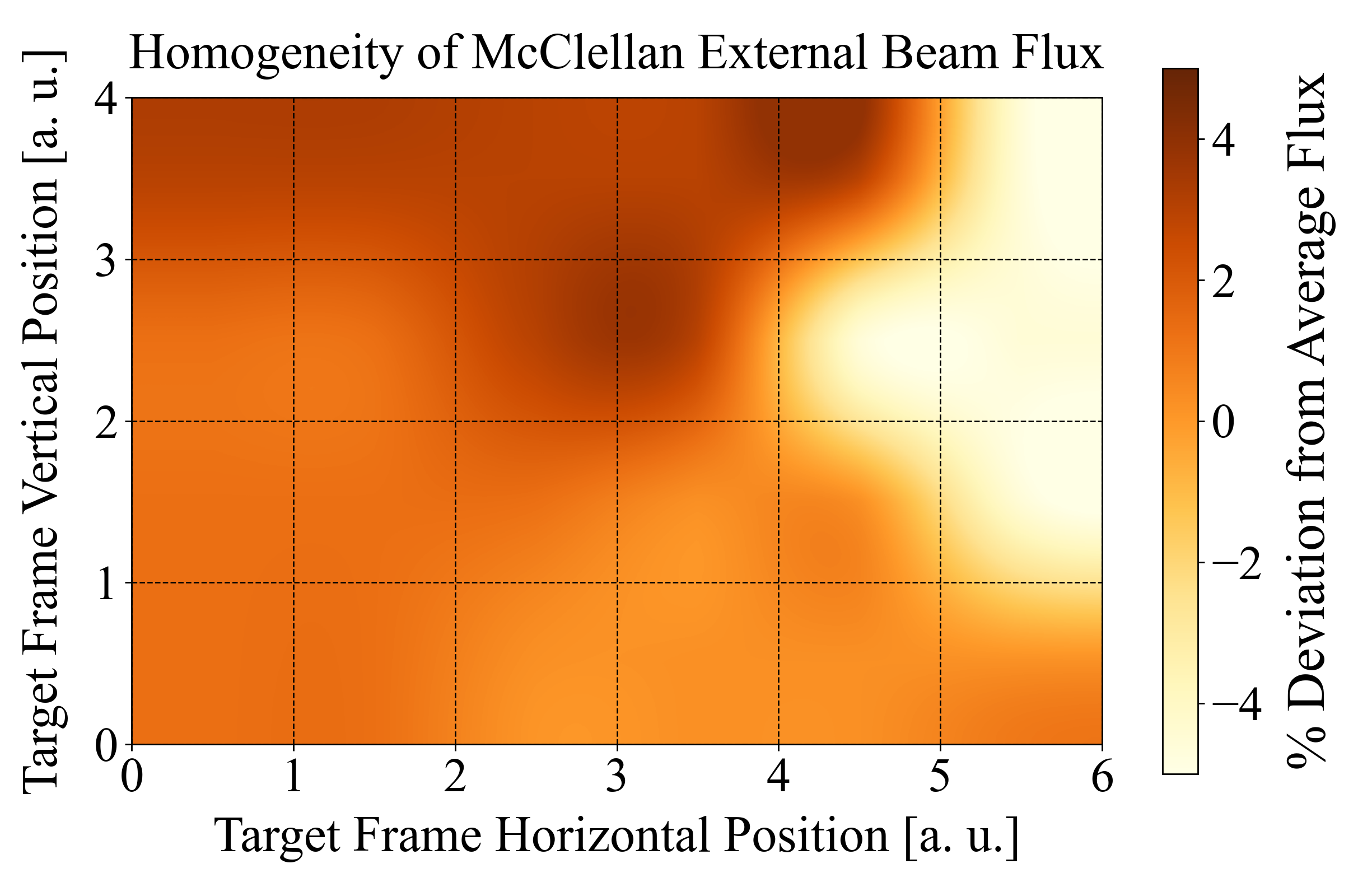}}}\hspace{6mm}
	\subfloat[Flux map across the irradiated target frame showing deviation from the average neutron flux at each position, as calculated from monitor foils. These results quantitatively describe the beam spot of the radiograph in Figure \ref{RadiographCrop}. The horizontal and vertical positions in this flux map correspond to the 4$\times$6 grid of targets, seen in Figure \ref{ShutterAndFrame}. \label{FluxMap}]
		{\includegraphics[width=0.65\textwidth, valign=c]{FluxMap}}
	\vspace{-0.15cm}
	\caption{Assessments of McClellan external bay beam uniformity at target irradiation site.}\label{UniformNeutronFlux}
\end{figure*}

Over the course of a year, numerous neutron irradiation experiments were conducted for the electroplated targets of interest to gather the required activation data and account for the particular McClellan beam conditions. In each neutron irradiation, the arsenic targets along with the monitor foils were mounted in a grid-like frame and fastened to the fast shutter of the McClellan bay, as illustrated in Figure \ref{ShutterAndFrame}. The fast shutter is typically used as means for ensuring uniform beam exposure in radiography experiments but it is additionally centered in the neutron beamline (see Figure \ref{McClellanBaySetup}) and faces an approximately $6.25\times 10^6$\,n/cm$^2\cdot$s thermal flux, which made it a useful mounting platform for the targets in this work. The flux impinging on the fast shutter is uniform across an approximately 15-inch diameter, which was preliminarily imaged by a radiograph, pictured in Figure \ref{RadiographCrop}, and later confirmed by the monitor foils (Figure \ref{FluxMap}). Irradiation times were $t_{irr}=5-8$\,hours and the targets were returned to LBNL for assessment within approximately 5 hours of the end-of-bombardment (EoB).

At LBNL, the neutron activation was measured through gamma spectroscopy using multiple ORTEC IDM-200-VTM HPGe detectors. The UC Berkeley code package Curie \cite{MorrellCurie}, with built-in nuclear structure and reaction databases, was used to analyze the gamma spectra. End-of-bombardment activities $A_0$ for the activation products of Table \ref{ActivationReactions} were determined from the count data with appropriate timing, efficiency, solid angle, and gamma attenuation corrections \cite{Morrell2020,Fox2020:NbLa}. The energy and absolute photopeak efficiency of the IDMs were calibrated using standard $^{57}$Co, $^{60}$Co, $^{133}$Ba, $^{137}$Cs, $^{152}$Eu, and $^{241}$Am sources. The efficiency model used in this work is a Curie-modified form of the semi-empirical formula proposed by \citet{Vidmar2001}.

\vspace{-0.2cm}
\begin{table*}[t]
\caption{Relevant nuclear data for neutron capture reactions used in the McClellan-based target characterization process \cite{DataSheetsA76,DataSheetsA198,DataSheetsA64,DataSheetsA59,ENDF,IRDFF2,Nyarko2010}. Uncertainties are listed in the least significant digit, that is, 1.314 (74) b means 1.314$\pm$0.074 b.}
\vspace{0.1cm}
\label{ActivationReactions}
\centering
\begin{tabular}{c|llll}
\hline\hline\\[-0.25cm]
Capture Reaction & Thermal Cross Section $\sigma_0$ [b] & Activation Product $t_{1/2}$ & Decay $E_\gamma$ [keV] & $I_\gamma$ [\%]\\[0.1cm]
\hline\\[-0.25cm]
$^{75}$As(n,$\gamma$)$^{76}$As & 4.28 (19) & 26.261 (17) h& 559.10 (5) &45.0 (2) \\[0.1cm]

$^{197}$Au(n,$\gamma$)$^{198}$Au & 98.70 (22) & 2.6941 (2) d & 411.80205 (17) &95.62 (6)  \\[0.1cm]

$^{63}$Cu(n,$\gamma$)$^{64}$Cu & 4.47 (18) & 12.701 (2) h & 1345.77 (6) &0.475 (11)  \\[0.1cm]

$^{58}$Fe(n,$\gamma$)$^{59}$Fe & 1.314 (74) & 44.490 (9) d & 1099.245 (3) &56.5 (9) \\[0.1cm]
\hline\hline
\end{tabular}
\vspace{-0.35cm}
\end{table*}

\subsubsection{Activation Analysis}
Given the measured count data, the initial calculation methodology for extracting the areal densities of the arsenic targets subsequent to neutron activation from the believed highly thermalized McClellan spectrum begins with \cite{Blaauw2017,Steinhauser2012,Karadag2003}
\begin{equation}
A_0=\phi N \sigma_0G_{th}g(1-e^{-\lambda t_{irr}}).
\end{equation}
Here, $\phi$ is the thermal neutron flux in the McClellan imaging bay, $N$ is the number of irradiated nuclei in the target under consideration, $\sigma_0$ is the appropriate thermal neutron capture cross section, $\lambda$ is the activation product decay constant, $G_{th}$ is the thermal neutron self-shielding factor for the irradiated target, and $g$ is the Westcott correction factor that accounts for the deviation of the capture cross section from a pure $1/v$ energy dependence. Typical $G_{th}$ corrective factors for micron-thick targets are in the 0.97--1.0 range and have established calculation methods \cite{Blaauw1995,Karadag2003,VanDo2008:W,VanDo2015:Sc}. From \citet{Mughabghab2003}, the Westcott correction factor for arsenic is 1.0005, for gold is 1.0054, for copper is 1.0002, and for iron is 1.0012.

Creating a relative activity measurement between the arsenic targets (subscript $As$) and a neutron monitor foil (subscript $mon$) yields
\begin{equation}\label{RelThermal}
\frac{A_{0,As}}{A_{0,mon}}=\frac{N_{As}\sigma_{0,As}G_{th,As}g_{As}(1-e^{-\lambda_{As} t_{irr}})}{N_{mon}\sigma_{0,mon}G_{th,mon}g_{mon}(1-e^{-\lambda_{mon} t_{irr}})}
\end{equation}
where there is no longer a dependence on neutron flux. The unknown arsenic electrodeposit masses, or equivalently the unknown number of arsenic nuclei in an electrodeposit, can be represented as
\begin{equation}
N_{As}=(\rho \Delta r)_{As} A_{plating} N_A /MM_{As},
\end{equation}
where $(\rho \Delta r)_{As}$ is the arsenic layer areal density, $A_{plating}$ is the electroplated area for the arsenic on the Ti backings, $N_A$ is Avogadro's number, and $MM_{As}$ is the molar mass of $^{75}$As. This expression is valid because of the earlier performed SEM work demonstrating the uniformity of the plated arsenic layers. Then, it follows that the desired arsenic target areal densities can be experimentally deduced from
\begin{equation}\label{ArealNaive}
\begin{split}
(\rho \Delta r)_{As}&=\frac{A_{0,As}N_{mon}\sigma_{0,mon}G_{th,mon}g_{mon}(1-e^{-\lambda_{mon} t_{irr}})}{A_{0,mon}\sigma_{0,As}G_{th,As}g_{As}(1-e^{-\lambda_{As} t_{irr}})}\\
& \times\frac{MM_{As}}{N_A A_{plating}}.
\end{split}
\end{equation}

However, early results using this methodology suggested that arsenic areal density magnitudes were 5--7$\times$ larger than expected based on the constraints set by limitations of the electroplating process and the electron transmission work. As well, the thermal neutron flux that could be extracted independently using the monitor foils, as was done in the flux map of Figure \ref{FluxMap}, was nearly a factor of five larger than the anticipated $6.25\times 10^6$\,n/cm$^2\cdot$s. These discrepancies were determined to be artifacts stemming from an as yet unaccounted for epithermal flux component in the McClellan reactor neutron spectrum.

Due to the fact the irradiation bay has historically only been used for thermal neutron imaging, the McClellan site has no descriptive epithermal flux data nor is there sufficient capture and elastic cross section knowledge of the materials under investigation to develop an analytical correction. In turn, it was necessary to perform a cadmium covered irradiation of the targets as one of the numerous experiments. Activation under cadmium sheets would remove the thermal neutron contribution to isolate the epithermal portion needed for activity subtraction as a corrective means for the inflated areal densities. Given the cadmium results, Equation (\ref{RelThermal}) is replaced by
\begin{equation}\label{RelCd}
\frac{R_{As}-F_{As,Cd}R_{As,Cd}}{R_{mon}-F_{mon,Cd}R_{mon,Cd}}=\frac{\sigma_{0,As}G_{th,As}g_{As}}{\sigma_{0,mon}G_{th,mon}g_{mon}},
\end{equation}
where $R_{As/mon}$ and $R_{As/mon,Cd}$ [Bq/target atom] are the reaction rates per atom for the targets after bare and Cd-covered irradiations, respectively, and $F_{As/mon,Cd}$ is the cadmium transmission correction factor. The monitor target reaction rates are directly calculated from measured $A_0$ and $N_{mon}$ values whereas $N_{As}$ and associated $(\rho \Delta r)_{As}$ remain the unknowns that are then solved for using a similar rearrangement to Equation (\ref{ArealNaive}).

However, these ensuing Cd-covered results likewise proved insufficient. It was seen that the approximate thermal-to-epithermal flux ratio ($f$) was $f\approx3$, indicating a very strong epithermal portion versus typical thermalized reactor fluxes that exist at $\approx f>20$ \cite{Chilian2008,Kumar2017,Dung2003}. Consequently, the predicted arsenic areal densities derived from the cadmium difference method were only slightly reduced versus the bare irradiations. Moreover, the relative subtraction in Equation ({\ref{RelCd}) was largely insensitive on account of the significant epithermal activation and yielded similar final ratios to Equation (\ref{RelThermal}). The failure of the well-established cadmium work to fully explain the 5--7$\times$ areal density overestimations pointed towards dominating missing corrections elsewhere in the activation calculations causing the apparent mass inflation.

Further analysis of the monitor foil data from the bare and cadmium experiments indicated that the convolution of target thickness and epithermal resonances was a sensitive parameter in the McClellan conditions and in fact played a significant role in the activation. Although the Cd-work should theoretically account for resonance effects and neutron flux perturbations associated with the non-thermal portion of the reactor spectrum, it is likely that the lack of detailed neutron flux shape and magnitude characterizations, either in-reactor or in the imaging bay at the beamline output, imparted an unknown influence on the experiments that obscured analysis.

To explore and root out these potential spectrum- and location-specific hindering factors would require a thorough reactor investigation that exists outside the aim of our electroplated target work and that can be in fact circumvented empirically in this application instead. Moreover, it is possible to experimentally derive an arsenic-specific calibration curve for the McClellan external beam conditions, which can provide a lump correction factor for combined thermal and epithermal self-shielding (absorption and scatter) effects without additional knowledge of the flux spectrum shape, flux perturbations from resonances in targets, or geometry contributions.

Typically, this type of overall effective self-shielding factor for a mixed neutron spectrum is composed from $G_{th}$ and its epithermal equivalent $G_{epi}$ and can be applied to bare target irradiation measurements \cite{Chilian2008}. Unfortunately, theoretical calculations for $G_{epi}$ are environment-dependent and contain inputs reliant on the current spectrum- and location-specific unknown factors at the McClellan site \cite{DeCorte1989,Yucel2004,DeCorte1979,Eastwood1962,Jovanovic1985,Karadag2003,Dung2003}. Therefore, our new empirical calibration approach aims to isolate and calculate this overall effective self-shielding factor when detailed flux characterizations not available. 

\subsubsection{Thick Pellet Calibration}
Accordingly, we produced thick pressed pellets of As$_2$O$_3$ of varying thickness for this calibration purpose where typical mass and dimensioning measurements could be made. Seven thick pressed pellets were created for this work, of 1.3\,cm diameter, via hydraulic press and trapezoidal split-sleeve dies with masses ranging from 100 to 900\,mg (see Figure \ref{PelletsOverview}). These pellets were irradiated under identical conditions at McClellan to the electroplated targets of interest and were likewise equivalently assayed by gamma spectroscopy at LBNL.

\vspace{-0.15cm}
\begin{figure}[H]
{\includegraphics[width=1.0\columnwidth]{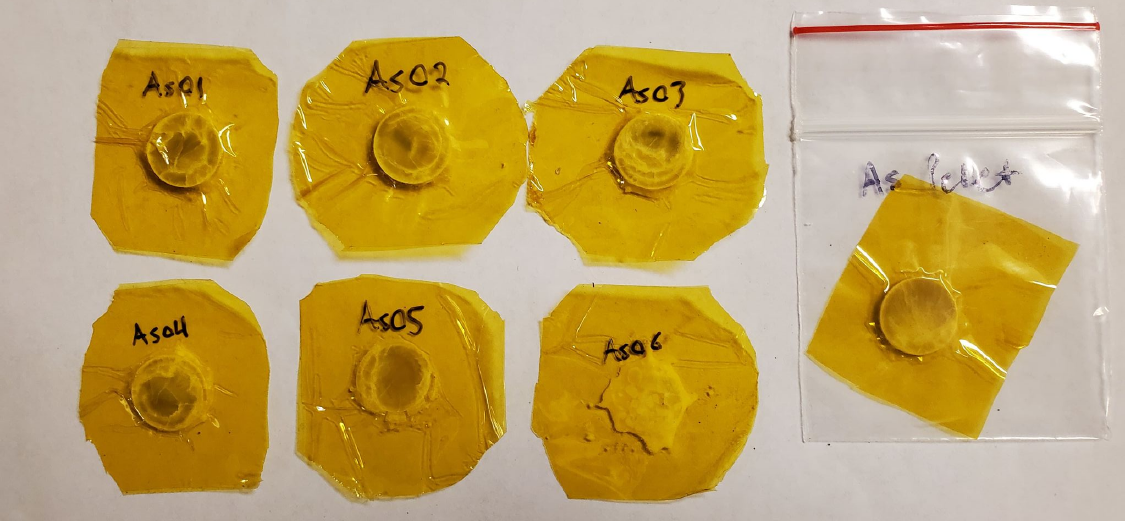}}
\vspace{-0.65cm}
\caption{Prepared thick pressed pellets of As$_2$O$_3$ of varying thickness used for calibration irradiations at McClellan. Note that although the disc structure of the ``As06" pellet could not be equally maintained during removal from the pressing process, its mass is known to equivalent uncertainty as all other pellets and its average diameter remains at 13\,mm.}
\label{PelletsOverview}
\end{figure}
\vspace{-0.3cm}

In contrast to the electroplated targets, however, the additional knowledge of mass and size for the thick pressed pellets mean that their induced $^{76}$As specific activities $\eta$ [Bq/atom] could be easily calculated from their determined $A_0$ values. These $\eta$ are valuable data points that were used to fit modified versions of theoretical self-shielding models as a function of mass/thickness. 

Specifically, the two accepted theoretical epithermal self-shielding corrective models of \citet{Karadag2003} and \citet{Martinho2003} were utilized for this purpose. They were modified from their original formulations in the sense that regression parameters have been added that account for not only epithermal shielding but also thermal self-shielding and all other non-independently described corrective factors for arsenic targets in the McClellan conditions. In this manner, the activation physics bases of these models are maintained but they are recast to provide general attenuation information.

In turn, this approach made it possible to extract the specific activity of a zero-thickness arsenic target in the McClellan beam, or more precisely, a Bq/atom production standard for $^{76}$As to directly apply to the measurements of the electroplated arsenic $A_0$. This calibration therefore simplifies the neutron activation areal density calculations and even renders the Au/Cu/Fe neutron monitor foil data moot.

\newcommand\ddfrac[2]{\frac{\displaystyle #1}{\displaystyle #2}}
Explicitly, the fit modified from the \citet{Karadag2003} $G_{epi}$ work is a function of the number of arsenic nuclei in a pellet per unit volume, $N_{pel}$, and pellet mean chord length, $d=2V/S\!A$ where $V$, $S\!A$ denote volume and total surface area of a pellet, respectively, as given in Equation (\ref{MustafaFit})
\begin{equation}\label{MustafaFit}
\eta = \eta_0\ddfrac{\left[I_v+\frac{\pi}{2}\sum_i\left(\frac{\Gamma_\gamma}{E_r}\right)_i\frac{\sigma(E_{ri})}{\sqrt{1+2N_{pel}\sigma(E_{ri})d}}\right]}{\left[I_v+\frac{\pi}{2}\sum_i\left(\frac{\Gamma_\gamma}{E_r}\right)_i\sigma(E_{ri})\right]},
\end{equation}
where $\eta_0$ is the zero-thickness specific activity fit parameter, $I_v$ is the fit parameter representing the $1/v$ contribution to the capture cross section resonance integral, $E_{ri}$ and $\Gamma_{\gamma i}$ are the energy and radiative width of the $i$th neutron resonance, and $\sigma(E_{ri})$ is the maximum total neutron cross section at that $i$th resonance. The resonance parameters of arsenic used for this work are presented in Table \ref{AsResonance} (see \ref{AppendixResonance}).

Similarly, from the \citet{Martinho2003} $G_{epi}$ approach, the adopted modified fit function is
\begin{equation}\label{UniversalCurve}
\eta = a\left[\frac{0.94}{1+\left(\ddfrac{z}{2.70}\right)^{0.82}}+b\right],
\end{equation}
where $a$ and $b$ are fit parameters and $z$ is a dimensionless effective target thickness defined by Equation (\ref{zDef}). The desired zero-thickness standard of activation quantity is found here by $\eta_0=\eta(z=0)$.
\begin{equation}\label{zDef}
z=1.5tN_{pel}\sum_i\sigma(E_{ri})\left(\frac{\Gamma_\gamma}{\Gamma}\right)^{0.5}_i,
\end{equation}
where $t$ is the measured pellet thickness and $\Gamma_i$ is the total width of the $i$th resonance.

Following the irradiation of the pellets at McClellan, and measurements of the accompanying $\eta$ at LBNL, the Equation (\ref{MustafaFit}) and (\ref{UniversalCurve}) fits were applied. The fit results of both models are shown in Figure \ref{Geff_Fitting_Plots} and the determined $\eta_0$ values are given in Table \ref{Eta0Values}. Both models agree within uncertainty on the zero-thickness standard of activation for arsenic at McClellan.

\begin{figure*}[t]
\centering
	\subfloat[Modified Karadag calibration fit form. \label{Geff_Fitting_Karadag}]
		{\includegraphics[width=0.5\textwidth]{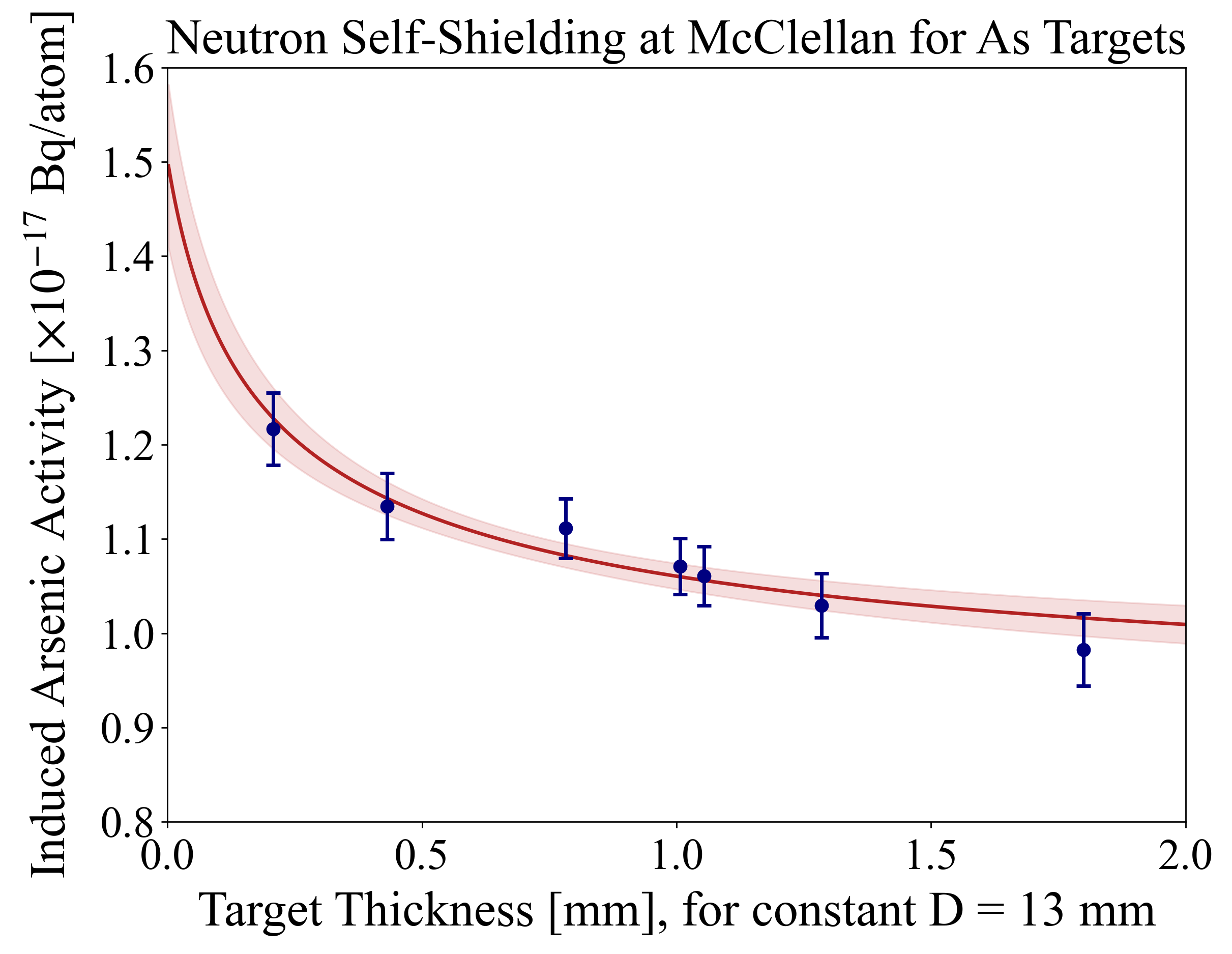}}
	\subfloat[Modified Martinho calibration fit form. \label{Geff_Fitting_Universal}]
		{\includegraphics[width=0.5\textwidth]{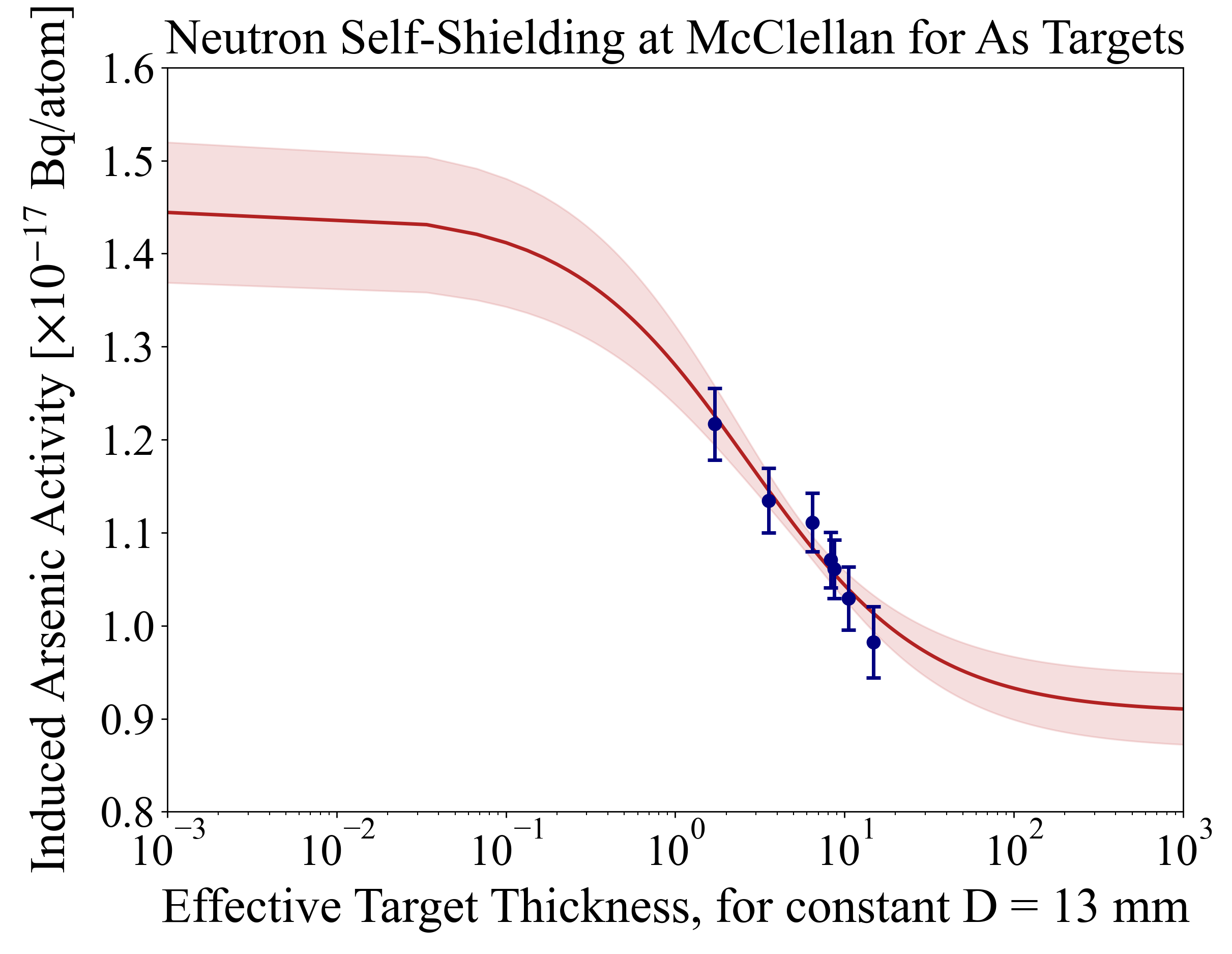}}
	\vspace{-0.15cm}
	\caption{Extraction of zero-thickness standard of activation for arsenic targets in McClellan external beam conditions by calibration curves using measured pellet $^{76}$As specific activity data. The bands surrounding the solid calibration curves represent fit errors at one standard deviation.}\label{Geff_Fitting_Plots}
\end{figure*}

\begin{table}[H]
\vspace{-0.21cm}
\caption{Zero-thickness standard of activation for arsenic targets in McClellan conditions as derived from the pellet calibration data.}
\vspace{0.1cm}
\label{Eta0Values}
\centering
\begin{tabular*}{\columnwidth}{l@{\extracolsep{\fill}}c}
\hline\hline\\[-0.25cm]
Calibration Fit Form & $\eta_0$ [Bq/atom] \\[0.1cm]
\hline\\[-0.25cm]
Modified Karadag    & (1.502 $\pm$ 0.086) $\times$10$^{-17}$ \\[0.1cm]
Modified Martinho    & (1.446 $\pm$ 0.076) $\times$10$^{-17}$\\[0.1cm]
\hline\hline
\end{tabular*}
\end{table}

Given the found $\eta_0$, the number of arsenic nuclei on each electroplated arsenic target of interest was very simply calculated using each target's measured EoB activity by
\begin{equation}\label{NewNumNucCalc}
N_{As}=\frac{A_{0,As}}{\eta_0}.
\end{equation}
The desired areal density quantity per electroplated target was finally directly calculated as
\begin{equation}\label{NewArealCalc}
(\rho\Delta r)_{As}=\frac{N_{As}MM_{As}}{N_A A_{plating}}.
\end{equation}

Note that an iterative procedure could have been applied to calculate individual $\eta$ values specific to each electroplated target. However, the fits do not have relevant precision at such small thickness values and the $\eta_0$ uncertainties instead contain any error from using the zero-thickness application to all targets equivalently. This choice to ubiquitously apply $\eta_0$ is further justified since the difference of $\eta$ versus $\eta_0$ for target thicknesses $<5$\,\si\um\ (see Table \ref{AsDensitiesTable}) is minimal regardless.

A variation of this effective corrective factor approach was performed by \citet{Chilian2008} for a larger variety of target nuclei. Their calculations still utilized measured condition-specific reactor parameters but it is clear that this overall methodology has established roots in literature and is viable. Origins for this technique in nuclear data experiments even reach farther back to \citet{Alfassi1982}, who performed thin foil mass measurements through neutron activation, though calculation details are not provided.

\section{Results and Discussion}
A summary of the produced electroplated targets for the stacked-target work and their characterization results by neutron activation with the determined $\eta_0$ are given in Table \ref{AsDensitiesTable}.

\begin{table*}[!t]
\caption{Properties of the 26 prepared arsenic targets by electrodeposition. The targets are separated into 3 groups according to their use in 3 different stacked-target irradiations and thus creation from different batches/electroplating experiments. Target thicknesses are average values calculated using the target areal densities determined through this work and a bulk density for arsenic of 5.72\,g/cm$^3$.}
\label{AsDensitiesTable}
\centering
\vspace{0.1cm}
\begin{tabular*}{\textwidth}{l@{\extracolsep{\fill}}lll}
\hline\hline\\[-0.25cm]
\multicolumn{4}{c}{\bfseries Target Stack \#1 (Two Electrode Cell)}\\[+0.1cm]
\hline\\[-0.25cm]
Target & Thickness [\si\um] & Areal Density from Activation [mg/cm$^2$] & Areal Density Uncertainty [\%]\\[+0.1cm]
\hline\\[-0.25cm]
As-SN1 &     3.24                   &  1.85                             &9.8 \\[0.1cm]

As-SN2 &        1.69                &       0.97              &9.9  \\[0.1cm]

As-SN3 &         1.81               &          1.04            &9.9  \\[0.1cm]

As-SN4 &        2.22                &          1.27             &10 \\[0.1cm]

As-SN5 &          1.95              &       1.12             &9.9 \\[0.1cm]

As-SN6 &           1.30             &        0.74             &11 \\[0.1cm]

As-SN7 &      2.36                  &         1.35          &8.9 \\[0.1cm]

As-SN8 &          0.94              &          0.54          &9.7 \\[0.1cm]

As-SN9 &            0.57            &           0.32            &10 \\[0.1cm]
\hline\\[-0.25cm]
\multicolumn{4}{c}{\bfseries Target Stack \#2 (Three Electrode Cell)}\\[+0.1cm]
\hline\\[-0.25cm]
Target & Thickness [\si\um] & Areal Density from Activation [mg/cm$^2$] & Areal Density Uncertainty [\%]\\[+0.1cm]
\hline\\[-0.25cm]
As-SN1 &      4.27                  &       2.45                 &8.2 \\[0.1cm]

As-SN2 &       4.30                 &          2.46              &8.3  \\[0.1cm]

As-SN3 &        3.62                &       2.07                 &9.0  \\[0.1cm]

As-SN4 &        3.54                &         2.03               &9.2 \\[0.1cm]

As-SN5 &          3.90              &       2.23               &8.7 \\[0.1cm]

As-SN6 &          3.11              &           1.78             &10\\[0.1cm]

As-SN7 &         2.79               &           1.59             &9.2 \\[0.1cm]

As-SN8 &            2.20            &           1.26             &9.0 \\[0.1cm]

As-SN9 &          2.57              &           1.47             &9.9 \\[0.1cm]

As-SN10 &          1.94            &         1.11                &10 \\[0.1cm]
\hline\\[-0.25cm]
\multicolumn{4}{c}{\bfseries Target Stack \#3 (Two Electrode Cell)}\\[+0.1cm]
\hline\\[-0.25cm]
Target & Thickness [\si\um] & Areal Density from Activation [mg/cm$^2$] & Areal Density Uncertainty [\%]\\[+0.1cm]
\hline\\[-0.25cm]
As-SN1 &        1.89               &      1.08                     &9.9 \\[0.1cm]

As-SN2 &        2.94               &       1.68                        &9.0  \\[0.1cm]

As-SN3 &         3.06              &       1.75                      &10  \\[0.1cm]

As-SN4 &        4.85               &          2.78                  &9.9 \\[0.1cm]

As-SN5 &        7.26               &        4.15                    &12 \\[0.1cm]

As-SN6 &        4.93               &         2.82            &9.0 \\[0.1cm]

As-SN7 &         12.6              &         7.22               &9.3 \\[0.1cm]
\hline\hline
\end{tabular*}
\end{table*}

The final electroplated target areal density uncertainties lie in the 8--10\% range, which can be contrasted with typical 0.1--1.0\% characterization uncertainties of commercially produced targets that are conventionally massed and measured. The uncertainties here have contributions from the $\eta_0$ calculations and pellet dimensional measurements, plating edge shape irregularities, and from the fitted peak areas, evaluated half-lives, gamma intensities, and detector efficiency calibrations used in the gamma spectroscopy measurements.

Uncertainties at the 8--10\% level will yield eventual cross section measurements with errors in the 9--15\% range. There is evidently room for improvement but this precision is still valid for stacked-target work because the experimental technique maps excitation functions over a range of energies and thus can inherently balance loss in precision with a cumulative result that still amounts to high quality data.

Furthermore, although these areal density uncertainties are large compared to commercially-sourced foils for other materials, it is important to note that pure arsenic targets, and particularly arsenic thin foils, have not been widely used to measure nuclear reaction data in the past. Most existing arsenic reaction data are from neutron-induced investigations and therefore have markedly different target requirements and measurement techniques. Very few charged-particle, and in turn stacked-target, pure arsenic foil measurements exist, meaning that the contribution from this work is valuable for extending knowledge in this context \cite{Mushtaq1988:ProtonsAs,Qaim1988:73SeIsomer}.

In this manner, paths forward from this work may include further more detailed electroplating studies. These studies exist outside of the applied nuclear data realm but would generate an improved description of underlying arsenic electrochemistry kinetics and consequences, likely deducing optimal cell assemblies and plating parameters, and perhaps leading to an arsenic target fabrication standardization. A fundamental study of this type has been performed by \citet{Wang2017:EutecticTargetFab}, where in-depth voltammetric work was used to explore arsenic electroplating charge-transfer kinetics, chronoamperometry was used to detail aspects of deposition nucleation and growth, and microstructure analysis was performed under different conditions. However, this study held a research focus within the semiconductor industry and was performed for a plating solution meeting associated needs. In turn, there is still demand for similarly conducted investigations for different plating solutions that also further examine substrate effects, deposition stress generation and impact, cell design, and consider precise characterization methods in the context of nuclear data experiments. A similar chemistry and materials science exploration is worthwhile with respect to arsenic vapor deposition as well. Adjacent considerations of fabrication through the creation and subsequent burning of arsine gas (AsH$_3$), akin to the Marsh Test \cite{MarshTest}, to yield a uniform arsenic mirror may too be worthwhile.

The implementation of neutron activation analysis in this work to extract areal densities rather than neutron reaction data or related structure parameters such as nuclear level densities is less common, but has evidently been shown as a viable target characterization approach. In cases of fragile targets, where appropriate neutron capture properties exist, this is an easily applied and accurate non-destructive technique. In combination with microstructure studies by SEM or similar, this characterization approach would fully describe fabricated targets.

Given the outcomes of this study, if this type of target assessment were to be adopted in future instances, it is clear that numerous improvements can be easily applied. Firstly, acquiring a more precise calibration curve would immediately reduce final uncertainties.

Likely, improved powder pressing technology that could make pellets which cover a broader range of thicknesses is a quickly attainable refinement. In fact, some powder pressing developments, such as the \citet{Sugai1997} and \citet{Esposito2019} vibrational/electrostatic work, even offer a pathway to the uniformity and thinness of electroplating experiments. However, it is useful to note that powder pressing pellets is also not a requirement for the activation calibration. Variations such as suspensions in solutions or casting with different compounds of the target of interest, where the secondary material does not offer competing neutron resonances, will all allow for equivalent experiments and derivations of a zero-thickness standard of activation. This foresight in the future means that dedicating more thought and care to this aspect of the experiment will certainly allow for a more precise calibration curve. It seems achievable to reach areal densities at reduced uncertainties near the theoretical minimum, defined by uncertainties of any used capture cross sections, generally in the 1--5\% range. Additional changes to experimental conditions such as moving to in-core activation for the targets to increase neutron flux and increasing irradiation time would improve this characterization approach. These changes would result in increased target activities, leading to more favourable gamma spectroscopy conditions that would reduce eventual contributing $A_0$ uncertainties.

Of course, if it is possible to use a well-characterized and well-thermalized reactor, the need for this calibration approach becomes optional as the cadmium difference method or equivalent techniques will work. However, this calibration approach may still be preferred since it requires only one experiment and relies on an extremely simple relative calculation where even monitor foils are unneeded. 

As an aside to target characterization, in our exploitation of neutron activation for this purpose, our fitting analyses have extracted what amounts to an integral strength measurement of resonance absorption and scatter. This easily conducted technique may find application in structure investigations or for integral data measurements relevant for the modeling of new reactor designs.

\section{Conclusions}
Thin uniform $^{75}$As targets on titanium backings appropriate for stacked-target experiments with areal densities from 0.32 to 7.2 mg/cm$^2$ were prepared by electrodeposition using an aqueous plating solution of As$_2$O$_3$ dissolved in HCl. Both two and three electrode electrolytic cells were used and best plating parameters were derived through voltammetric investigations. Significant efforts were devoted to target characterization, which included surface morphology studies by SEM and non-destructive thickness assays by neutron activation. The applied neutron activation analysis was formulated to be independent of neutron flux shape, environmental factors, and source geometry, while correcting for any potential scatter or absorption effects.

\section*{Acknowledgments}
This research was supported by the U.S. Department of Energy Isotope Program, managed by the Office of Science for Isotope R\&D and Production, and was carried out under Lawrence Berkeley National Laboratory (Contract No. DE-AC02-05CH11231), Los Alamos National Laboratory (Contract No. 89233218CNA000001), and Brookhaven National Laboratory (Contract No. DEAC02-98CH10886). The authors acknowledge the assistance and support of the operations and facilities staff of the McClellan Nuclear Research Center. We thank James Seals and the rest of the LBNL Radiation Protection Group. The authors also acknowledge the Biomolecular Nanotechnology Center's (BNC) facilities at UC Berkeley.
\clearpage

\onecolumn
\appendix
\section{Resonance Data for Neutron Self-Shielding Calculations}\label{AppendixResonance}
The properties of neutron resonances in $^{75}$As relevant to the epithermal calculations at the McClellan site are listed in Table \ref{AsResonance}.
\begin{table*}[!htb]
\caption{Resonance parameters used in self-shielding calculations of arsenic pellets \cite{Karadag2003,Sukhoruchkin2015,Mughabghab1973,ENDF}.}
\label{AsResonance}
\centering
\vspace{0.1cm}
\begin{tabular}{p{0.225\textwidth}p{0.225\textwidth}p{0.225\textwidth}p{0.225\textwidth}}
\hline\hline\\[-0.25cm]
\multicolumn{4}{c}{\bfseries $^{75}$As Neutron Resonance Parameters}\\[+0.1cm]
\hline\\[-0.25cm]
\multirow{2}{=}{Resonance Energy $E_r$ [eV]}&\multirow{2}{=}{Radiative Width of the Resonance $\Gamma_\gamma$ [eV]}&\multirow{2}{=}{Total Width of the Resonance $\Gamma$ [eV] }&\multirow{2}{=}{Cross Section at $E_r$ [b]}\\ \\[0.1cm]
\hline\\[-0.25cm]
47.00 & 0.26 & 0.29  & 2544.1                      \\[0.1cm]

92.40 & 0.25 &  0.28 &  394.4                  \\[0.1cm]

252.7 & 0.27 & 0.32   &   349.4          \\[0.1cm]

318.6 & 0.30 & 0.77   &  1909.7        \\[0.1cm]

326.7 & 0.35 &  0.89   & 84.6   \\[0.1cm]

455.5 & 0.34 & 0.37  & 1205.6  \\[0.1cm]

493.3 & 0.30 & 0.35 & 300.532                  \\[0.1cm]

533.4 & 0.28 & 0.30  & 2507.0               \\[0.1cm]

664.9 & 0.30 & 0.62 & 411.375              \\[0.1cm]

733.9 & 0.35 & 1.5 & 802.415                 \\[0.1cm]

737.4 & 0.30 & 2.3 & 1608.4535              \\[0.1cm]
\hline\hline
\end{tabular}
\end{table*}
\clearpage

\twocolumn
\biboptions{sort&compress}
\bibliographystyle{elsarticle-num-names_etal} 

\begin{thebibliography}{62}
\expandafter\ifx\csname natexlab\endcsname\relax\def\natexlab#1{#1}\fi
\providecommand{\url}[1]{\texttt{#1}}
\providecommand{\href}[2]{#2}
\providecommand{\path}[1]{#1}
\providecommand{\DOIprefix}{doi:}
\providecommand{\ArXivprefix}{arXiv:}
\providecommand{\URLprefix}{URL: }
\providecommand{\Pubmedprefix}{pmid:}
\providecommand{\doi}[1]{\href{http://dx.doi.org/#1}{\path{#1}}}
\providecommand{\Pubmed}[1]{\href{pmid:#1}{\path{#1}}}
\providecommand{\bibinfo}[2]{#2}
\ifx\xfnm\relax \def\xfnm[#1]{\unskip,\space#1}\fi
\bibitem[{Fox \textit{et~al.}(2021)Fox, Voyles, Morrell, Bernstein, Batchelder,
  Birnbaum, Cutler, Koning, Lewis, Medvedev, Nortier, O'Brien, and
  Vermeulen}]{Fox2021:As}
\bibinfo{author}{M.~B. Fox}, \bibinfo{author}{A.~S. Voyles},
  \bibinfo{author}{J.~T. Morrell}, \bibinfo{author}{L.~A. Bernstein},
  \bibinfo{author}{J.~C. Batchelder}, \bibinfo{author}{E.~R. Birnbaum},
  \bibinfo{author}{C.~S. Cutler}, \bibinfo{author}{A.~J. Koning},
  \bibinfo{author}{A.~M. Lewis}, \bibinfo{author}{D.~G. Medvedev},
  \bibinfo{author}{F.~M. Nortier}, \bibinfo{author}{E.~M. O'Brien},
  \bibinfo{author}{C.~Vermeulen},
\newblock \bibinfo{title}{Measurement and modeling of proton-induced reactions
  on arsenic from 35 to 200 mev},
\newblock \bibinfo{journal}{Physical Review C} \bibinfo{volume}{104}
  (\bibinfo{year}{2021}) \bibinfo{pages}{064615}.
  \DOIprefix\doi{10.1103/PhysRevC.104.064615}.
\bibitem[{Wang \textit{et~al.}(2017)Wang, Hsieh, and
  Sun}]{Wang2017:EutecticTargetFab}
\bibinfo{author}{P.~Wang}, \bibinfo{author}{Y.~Hsieh},
  \bibinfo{author}{I.~Sun},
\newblock \bibinfo{title}{{On the Electrodeposition of Arsenic in a Choline
  Chloride/Ethylene Glycol Deep Eutectic Solvent}},
\newblock \bibinfo{journal}{Journal of The Electrochemical Society}
  \bibinfo{volume}{164} (\bibinfo{year}{2017}) \bibinfo{pages}{D204--D209}.
  \DOIprefix\doi{10.1149/2.1061704jes}.
\bibitem[{Chandra \textit{et~al.}(1988)Chandra, Khare, and
  Upadhyaya}]{Chandra1988}
\bibinfo{author}{S.~Chandra}, \bibinfo{author}{N.~Khare},
  \bibinfo{author}{H.~M. Upadhyaya},
\newblock \bibinfo{title}{{Photoelectrochemical solar cells using
  electrodeposited GaAs and AlSb semiconductor films}},
\newblock \bibinfo{journal}{Bulletin of Materials Science} \bibinfo{volume}{10}
  (\bibinfo{year}{1988}) \bibinfo{pages}{323--332}.
  \DOIprefix\doi{10.1007/BF02744303}.
\bibitem[{Rohm and Munzel(1972)}]{Rohm1972}
\bibinfo{author}{H.~F. Rohm}, \bibinfo{author}{H.~Munzel},
\newblock \bibinfo{title}{{Excitation Functions for Deuteron Reactions with
  $^{75}$As}},
\newblock \bibinfo{journal}{Journal of Inorganic and Nuclear Chemistry}
  \bibinfo{volume}{34} (\bibinfo{year}{1972}) \bibinfo{pages}{1773--1784}.
  \DOIprefix\doi{10.1016/0022-1902(72)80523-2}.
\bibitem[{Voyles(2020)}]{Voyles2020:WANDA}
\bibinfo{author}{A.~S. Voyles},
\newblock \bibinfo{title}{{Targetry Fabrication for Nuclear Data
  Measurements}},
\newblock in: \bibinfo{booktitle}{Workshop for Applied Nuclear Data
  Activities}, \bibinfo{address}{Washington, D.C.}, \bibinfo{year}{2020}.
  \URLprefix \url{https://conferences.lbl.gov/event/292/}.
\bibitem[{Qaim \textit{et~al.}(1986)Qaim, Blessing, and
  Ollig}]{Qaim1986:AsTarg}
\bibinfo{author}{S.~M. Qaim}, \bibinfo{author}{G.~Blessing},
  \bibinfo{author}{H.~Ollig},
\newblock \bibinfo{title}{{Excitation Functions of
  $^{75}$As($\alpha$,n)$^{78}$Br and $^{75}$As($\alpha$,2n)$^{77m,g}$Br
  Reactions from Threshold to 28 MeV}},
\newblock \bibinfo{journal}{Radiochimica Acta} \bibinfo{volume}{39}
  (\bibinfo{year}{1986}) \bibinfo{pages}{57--60}.
  \DOIprefix\doi{10.1524/ract.1986.39.2.57}.
\bibitem[{Fowler \textit{et~al.}(2007)Fowler, {J. Chou}, Jones, and
  Chen}]{ToxicHandbook}
\bibinfo{author}{B.~A. Fowler}, \bibinfo{author}{C.~S. {J. Chou}},
  \bibinfo{author}{R.~L. Jones}, \bibinfo{author}{C.~Chen},
\newblock \bibinfo{title}{{Chapter 19 - Arsenic}},
\newblock in: \bibinfo{editor}{G.~F. Nordberg}, \bibinfo{editor}{B.~A. Fowler},
  \bibinfo{editor}{M.~Nordberg}, \bibinfo{editor}{L.~T. Friberg} (Eds.),
  \bibinfo{booktitle}{Handbook on the Toxicology of Metals},
  \bibinfo{edition}{third} ed., \bibinfo{publisher}{Academic Press},
  \bibinfo{year}{2007}, pp. \bibinfo{pages}{367--406}.
  \DOIprefix\doi{10.1016/B978-012369413-3/50074-4}.
\bibitem[{{Alfa Aesar}(2021)}]{AlfaAesar}
\bibinfo{author}{{Alfa Aesar}}, \bibinfo{title}{{Arsenic Foil Products}},
  \bibinfo{year}{2021}. \URLprefix
  \url{https://www.alfa.com/en/search/?q=arsenic+foil}.
\bibitem[{Goodfellow(2021)}]{Goodfellow}
\bibinfo{author}{Goodfellow}, \bibinfo{title}{{Arsenic Product Search}},
  \bibinfo{year}{2021}. \URLprefix
  \url{http://www.goodfellow.com/catalogue/GFCat4.php?ewd_token=UiV5M4AHpBAqPJii1ozeO1Tzq0koaE&n=VL5x23JEGfW6jv6polbD1iNwYNHr9c}.
\bibitem[{Breunig \textit{et~al.}(2017)Breunig, Spahn, Hermanne, Spellerberg,
  Scholten, and Coenen}]{Breunig2017}
\bibinfo{author}{K.~Breunig}, \bibinfo{author}{I.~Spahn},
  \bibinfo{author}{A.~Hermanne}, \bibinfo{author}{S.~Spellerberg},
  \bibinfo{author}{B.~Scholten}, \bibinfo{author}{H.~H. Coenen},
\newblock \bibinfo{title}{{Cross section measurements of
  $^{75}$As($\alpha$,xn)$^{76,77,78}$Br and $^{75}$As($\alpha$,x)$^{74}$As
  nuclear reactions using the monitor radionuclides $^{67}$Ga and $^{66}$Ga for
  beam evaluation}},
\newblock \bibinfo{journal}{Radiochimica Acta} \bibinfo{volume}{105}
  (\bibinfo{year}{2017}) \bibinfo{pages}{431--439}.
  \DOIprefix\doi{10.1515/ract-2016-2593}.
\bibitem[{Nozaki \textit{et~al.}(1979)Nozaki, Iwamoto, and Itoh}]{Nozaki1979}
\bibinfo{author}{T.~Nozaki}, \bibinfo{author}{M.~Iwamoto},
  \bibinfo{author}{Y.~Itoh},
\newblock \bibinfo{title}{{Production of $^{77}$Br by various nuclear
  reactions}},
\newblock \bibinfo{journal}{The International Journal Of Applied Radiation And
  Isotopes} \bibinfo{volume}{30} (\bibinfo{year}{1979})
  \bibinfo{pages}{79--83}. \DOIprefix\doi{10.1016/0020-708X(79)90137-6}.
\bibitem[{Paans \textit{et~al.}(1980)Paans, Welleweerd, Vaalburg, Reiffers, and
  Woldring}]{Paans1980}
\bibinfo{author}{A.~M. Paans}, \bibinfo{author}{J.~Welleweerd},
  \bibinfo{author}{W.~Vaalburg}, \bibinfo{author}{S.~Reiffers},
  \bibinfo{author}{M.~G. Woldring},
\newblock \bibinfo{title}{{Excitation functions for the production of
  bromine-75: A potential nuclide for the labelling of radiopharmaceuticals}},
\newblock \bibinfo{journal}{The International Journal Of Applied Radiation And
  Isotopes} \bibinfo{volume}{31} (\bibinfo{year}{1980})
  \bibinfo{pages}{267--273}. \DOIprefix\doi{10.1016/0020-708X(80)90032-0}.
\bibitem[{Alfassi and Weinreich(1982)}]{Alfassi1982}
\bibinfo{author}{Z.~Alfassi}, \bibinfo{author}{R.~Weinreich},
\newblock \bibinfo{title}{{The Production of Positron Emitters $^{75}$Br and
  $^{76}$Br: Excitation Functions and Yields for $^3$He and $\alpha$-Particle
  Induced Nuclear Reactions on Arsenic}},
\newblock \bibinfo{journal}{Radiochimica Acta} \bibinfo{volume}{30}
  (\bibinfo{year}{1982}) \bibinfo{pages}{67--71}.
  \DOIprefix\doi{10.1524/ract.1982.30.2.67}.
\bibitem[{Waters \textit{et~al.}(1973)Waters, Nunn, and Thakur}]{Waters1973}
\bibinfo{author}{S.~L. Waters}, \bibinfo{author}{A.~D. Nunn},
  \bibinfo{author}{M.~L. Thakur},
\newblock \bibinfo{title}{{Cross-Section Measurements for the
  $^{75}$As($\alpha$,2n)$^{77}$Br Reaction}},
\newblock \bibinfo{journal}{Journal of Inorganic Nuclear Chemistry}
  \bibinfo{volume}{35} (\bibinfo{year}{1973}) \bibinfo{pages}{3413--3416}.
  \DOIprefix\doi{10.1016/0022-1902(73)80346-X}.
\bibitem[{Brodovitch \textit{et~al.}(1976)Brodovitch, Hogan, and
  Burns}]{Brodovitch1976:ProtonsPEM}
\bibinfo{author}{J.~C. Brodovitch}, \bibinfo{author}{J.~J. Hogan},
  \bibinfo{author}{K.~Burns},
\newblock \bibinfo{title}{{The Pre-equilibrium Statistical Model: Comparison of
  Calculations With Two (p,xn) Reactions}},
\newblock \bibinfo{journal}{Journal of Inorganic and Nuclear Chemistry}
  \bibinfo{volume}{38} (\bibinfo{year}{1976}) \bibinfo{pages}{1581--1586}.
  \DOIprefix\doi{10.1016/0022-1902(76)80639-2}.
\bibitem[{Morrison and {Caretto Jr.}(1962)}]{Morrison1962}
\bibinfo{author}{D.~Morrison}, \bibinfo{author}{A.~{Caretto Jr.}},
\newblock \bibinfo{title}{{Excitation Functions of (p,xp) Reactions}},
\newblock \bibinfo{journal}{Physical Review} \bibinfo{volume}{127}
  (\bibinfo{year}{1962}) \bibinfo{pages}{1731}.
  \DOIprefix\doi{10.1103/PhysRev.127.1731}.
\bibitem[{Johnson \textit{et~al.}(1958)Johnson, Galonsky, and
  Ulrich}]{Johnson1958}
\bibinfo{author}{C.~H. Johnson}, \bibinfo{author}{A.~Galonsky},
  \bibinfo{author}{J.~P. Ulrich},
\newblock \bibinfo{title}{{Proton strength functions from (p,n) cross
  sections}},
\newblock \bibinfo{journal}{Physical Review} \bibinfo{volume}{109}
  (\bibinfo{year}{1958}) \bibinfo{pages}{1243--1254}.
  \DOIprefix\doi{10.1103/PhysRev.109.1243}.
\bibitem[{Bowen and {Irvine Jr.}(1962)}]{Bowen1962}
\bibinfo{author}{L.~Bowen}, \bibinfo{author}{J.~{Irvine Jr.}},
\newblock \bibinfo{title}{{Nuclear Excitation Functions and Thick-Target Yields
  F$^{19}$, Na$^{23}$, As$^{75}$(d,t), and Na$^{23}$, As$^{75}$(d,p)}},
\newblock \bibinfo{journal}{Physical Review} \bibinfo{volume}{127}
  (\bibinfo{year}{1962}) \bibinfo{pages}{1698}.
  \DOIprefix\doi{10.1103/PhysRev.127.1698}.
\bibitem[{Mushtaq \textit{et~al.}(1988)Mushtaq, Qaim, and
  St{\"{o}}cklin}]{Mushtaq1988:ProtonsAs}
\bibinfo{author}{A.~Mushtaq}, \bibinfo{author}{S.~M. Qaim},
  \bibinfo{author}{G.~St{\"{o}}cklin},
\newblock \bibinfo{title}{{Production of $^{73}$Se via (p, 3n) and (d, 4n)
  Reactions on Arsenic}},
\newblock \bibinfo{journal}{International Journal of Radiation Applications and
  Instrumentation Part A} \bibinfo{volume}{39} (\bibinfo{year}{1988})
  \bibinfo{pages}{1085--1091}. \DOIprefix\doi{10.1016/0883-2889(88)90146-3}.
\bibitem[{Information(2008)}]{AsOxideTemp}
\bibinfo{author}{I.~P. R. C.~S. Information}, \bibinfo{title}{Arsenic Trioxide,
  International Chemical Safety Card: 0378},
  \bibinfo{organization}{International Programme on Chemical Safety, World
  Health Organization}, \bibinfo{year}{2008}. \URLprefix
  \url{https://inchem.org/documents/icsc/icsc/eics0378.htm},
  \bibinfo{note}{uRL: https://inchem.org/documents/icsc/icsc/eics0378.htm}.
\bibitem[{Fox \textit{et~al.}(2021)Fox, Voyles, Morrell, Bernstein, Lewis,
  Koning, Batchelder, Birnbaum, Cutler, Medvedev, Nortier, O'Brien, and
  Vermeulen}]{Fox2020:NbLa}
\bibinfo{author}{M.~B. Fox}, \bibinfo{author}{A.~S. Voyles},
  \bibinfo{author}{J.~T. Morrell}, \bibinfo{author}{L.~A. Bernstein},
  \bibinfo{author}{A.~M. Lewis}, \bibinfo{author}{A.~J. Koning},
  \bibinfo{author}{J.~C. Batchelder}, \bibinfo{author}{E.~R. Birnbaum},
  \bibinfo{author}{C.~S. Cutler}, \bibinfo{author}{D.~G. Medvedev},
  \bibinfo{author}{F.~M. Nortier}, \bibinfo{author}{E.~M. O'Brien},
  \bibinfo{author}{C.~Vermeulen},
\newblock \bibinfo{title}{{Investigating high-energy proton-induced reactions
  on spherical nuclei: Implications for the preequilibrium exciton model}},
\newblock \bibinfo{journal}{Physical Review C} \bibinfo{volume}{103}
  (\bibinfo{year}{2021}) \bibinfo{pages}{034601}.
  \DOIprefix\doi{10.1103/PhysRevC.103.034601}.
\bibitem[{Voyles \textit{et~al.}(2018)Voyles, Bernstein, Birnbaum, Engle,
  Graves, Kawano, Lewis, and Nortier}]{Voyles2018:Nb}
\bibinfo{author}{A.~S. Voyles}, \bibinfo{author}{L.~A. Bernstein},
  \bibinfo{author}{E.~R. Birnbaum}, \bibinfo{author}{J.~W. Engle},
  \bibinfo{author}{S.~A. Graves}, \bibinfo{author}{T.~Kawano},
  \bibinfo{author}{A.~M. Lewis}, \bibinfo{author}{F.~M. Nortier},
\newblock \bibinfo{title}{{Excitation functions for (p,x) reactions of niobium
  in the energy range of E$_p=40-90$ MeV}},
\newblock \bibinfo{journal}{Nuclear Instruments and Methods in Physics Research
  B} \bibinfo{volume}{429} (\bibinfo{year}{2018}) \bibinfo{pages}{53--74}.
  \DOIprefix\doi{10.1016/j.nimb.2018.05.028}.
\bibitem[{Marus \textit{et~al.}(2015)Marus, Engle, John, Birnbaum, and
  Nortier}]{Marus2015}
\bibinfo{author}{L.~A. Marus}, \bibinfo{author}{J.~W. Engle},
  \bibinfo{author}{K.~D. John}, \bibinfo{author}{E.~R. Birnbaum},
  \bibinfo{author}{F.~M. Nortier},
\newblock \bibinfo{title}{{Experimental and computational techniques for the
  analysis of proton beam propagation through a target stack}},
\newblock \bibinfo{journal}{Nuclear Instruments and Methods in Physics Research
  B} \bibinfo{volume}{345} (\bibinfo{year}{2015}) \bibinfo{pages}{48--52}.
  \DOIprefix\doi{10.1016/j.nimb.2014.12.048}.
\bibitem[{Graves \textit{et~al.}(2016)Graves, Ellison, Barnhart, Valdovinos,
  Eva, Nortier, Nickles, and Engle}]{Graves2016:StackTarget}
\bibinfo{author}{S.~A. Graves}, \bibinfo{author}{P.~A. Ellison},
  \bibinfo{author}{T.~E. Barnhart}, \bibinfo{author}{H.~F. Valdovinos},
  \bibinfo{author}{R.~Eva}, \bibinfo{author}{F.~M. Nortier},
  \bibinfo{author}{R.~J. Nickles}, \bibinfo{author}{J.~W. Engle},
\newblock \bibinfo{title}{{Nuclear excitation functions of proton-induced
  reactions (E$_p=35-90$ MeV) from Fe, Cu, Al}},
\newblock \bibinfo{journal}{Nuclear Instruments and Methods in Physics Research
  B} \bibinfo{volume}{386} (\bibinfo{year}{2016}) \bibinfo{pages}{44--53}.
  \DOIprefix\doi{10.1016/j.nimb.2016.09.018.Nuclear}.
\bibitem[{Seidl \textit{et~al.}(2019)Seidl, Balázs, and Scheer}]{Seidl2019}
\bibinfo{author}{M.~Seidl}, \bibinfo{author}{G.~Balázs},
  \bibinfo{author}{M.~Scheer},
\newblock \bibinfo{title}{The chemistry of yellow arsenic},
\newblock \bibinfo{journal}{Chemical Reviews} \bibinfo{volume}{119}
  (\bibinfo{year}{2019}) \bibinfo{pages}{8406--8434}.
  \DOIprefix\doi{10.1021/acs.chemrev.8b00713}.
\bibitem[{Fassbender \textit{et~al.}(2009)Fassbender, Bach, Bond, Nortier, and
  Vieira}]{Fassbender2009:TargFab}
\bibinfo{author}{M.~Fassbender}, \bibinfo{author}{H.~Bach},
  \bibinfo{author}{E.~Bond}, \bibinfo{author}{F.~M. Nortier},
  \bibinfo{author}{D.~Vieira},
\newblock \bibinfo{title}{{Preparation of thin arsenic and radioarsenic targets
  for neutron capture studies}},
\newblock \bibinfo{journal}{Journal of Radioanalytical and Nuclear Chemistry}
  \bibinfo{volume}{282} (\bibinfo{year}{2009}) \bibinfo{pages}{365--368}.
  \DOIprefix\doi{10.1007/s10967-009-0145-0}.
\bibitem[{Menzies and Owen(1966)}]{Menzies1966}
\bibinfo{author}{I.~A. Menzies}, \bibinfo{author}{L.~W. Owen},
\newblock \bibinfo{title}{{The electrodeposition of arsenic from aqueous and
  non-aqueous solutions}},
\newblock \bibinfo{journal}{Electrochimica Acta} \bibinfo{volume}{11}
  (\bibinfo{year}{1966}) \bibinfo{pages}{251--265}.
  \DOIprefix\doi{10.1016/0013-4686(66)80012-9}.
\bibitem[{Lupinacci \textit{et~al.}(2014)Lupinacci, Kacher, Eilenberg, Shapiro,
  Hosemann, and Minor}]{Lupinacci2014}
\bibinfo{author}{A.~Lupinacci}, \bibinfo{author}{J.~Kacher},
  \bibinfo{author}{A.~Eilenberg}, \bibinfo{author}{A.~A. Shapiro},
  \bibinfo{author}{P.~Hosemann}, \bibinfo{author}{A.~M. Minor},
\newblock \bibinfo{title}{{Cryogenic in situ microcompression testing of Sn}},
\newblock \bibinfo{journal}{Acta Materialia} \bibinfo{volume}{78}
  (\bibinfo{year}{2014}) \bibinfo{pages}{56--64}.
  \DOIprefix\doi{10.1016/j.actamat.2014.06.026}.
\bibitem[{Frazer \textit{et~al.}(2015)Frazer, Abad, Krumwiede, Back, Khalifa,
  Deck, and Hosemann}]{Frazer2015}
\bibinfo{author}{D.~Frazer}, \bibinfo{author}{M.~D. Abad},
  \bibinfo{author}{D.~Krumwiede}, \bibinfo{author}{C.~A. Back},
  \bibinfo{author}{H.~E. Khalifa}, \bibinfo{author}{C.~P. Deck},
  \bibinfo{author}{P.~Hosemann},
\newblock \bibinfo{title}{{Localized mechanical property assessment of SiC/SiC
  composite materials}},
\newblock \bibinfo{journal}{Composites: Part A} \bibinfo{volume}{70}
  (\bibinfo{year}{2015}) \bibinfo{pages}{93--101}.
  \DOIprefix\doi{10.1016/j.compositesa.2014.11.008}.
\bibitem[{Sugai(1997)}]{Sugai1997}
\bibinfo{author}{I.~Sugai},
\newblock \bibinfo{title}{{An application of a new type deposition method to
  nuclear target preparation}},
\newblock \bibinfo{journal}{Nuclear Instruments and Methods in Physics Research
  A} \bibinfo{volume}{397} (\bibinfo{year}{1997}) \bibinfo{pages}{81--90}.
  \DOIprefix\doi{10.1016/S0168-9002(97)00733-X}.
\bibitem[{{Research Reactor Safety Analysis Services}(2000)}]{McClellanNRC}
\bibinfo{author}{{Research Reactor Safety Analysis Services}},
  \bibinfo{title}{{McClellan Nuclear Radiation Center Safety Analysis Report
  Revision 4, License No. R-130, Docket No. 50-607}}, \bibinfo{type}{Technical
  Report}, University of California, Davis, \bibinfo{address}{Sacramento,
  California}, \bibinfo{year}{2000}.
\bibitem[{Morrell(2020)}]{MorrellCurie}
\bibinfo{author}{J.~T. Morrell}, \bibinfo{title}{{Curie: Python Toolkit for
  Experimental Nuclear Data}}, \bibinfo{year}{2020}. \URLprefix
  \url{https://pypi.org/project/curie/}.
\bibitem[{Morrell \textit{et~al.}(2020)Morrell, Voyles, Basunia, Batchelder,
  Matthews, and Bernstein}]{Morrell2020}
\bibinfo{author}{J.~T. Morrell}, \bibinfo{author}{A.~S. Voyles},
  \bibinfo{author}{M.~S. Basunia}, \bibinfo{author}{J.~C. Batchelder},
  \bibinfo{author}{E.~F. Matthews}, \bibinfo{author}{L.~A. Bernstein},
\newblock \bibinfo{title}{{Measurement of $^{139}$La(p,x) cross sections from
  35-60 MeV by stacked-target activation}},
\newblock \bibinfo{journal}{The European Physical Journal A}
  \bibinfo{volume}{56} (\bibinfo{year}{2020}) \bibinfo{pages}{13}.
  \DOIprefix\doi{10.1140/epja/s10050-019-00010-0}.
\bibitem[{Vidmar \textit{et~al.}(2001)Vidmar, Korun, Likar, and
  Martin{\v{c}}i{\v{c}}}]{Vidmar2001}
\bibinfo{author}{T.~Vidmar}, \bibinfo{author}{M.~Korun},
  \bibinfo{author}{A.~Likar}, \bibinfo{author}{R.~Martin{\v{c}}i{\v{c}}},
\newblock \bibinfo{title}{{A semi-empirical model of the efficiency curve for
  extended sources in gamma-ray spectrometry}},
\newblock \bibinfo{journal}{Nuclear Instruments and Methods in Physics Research
  A} \bibinfo{volume}{470} (\bibinfo{year}{2001}) \bibinfo{pages}{533--547}.
  \DOIprefix\doi{10.1016/S0168-9002(01)00799-9}.
\bibitem[{Singh(1995)}]{DataSheetsA76}
\bibinfo{author}{B.~Singh},
\newblock \bibinfo{title}{{Nuclear Data Sheets Update for A = 76}},
\newblock \bibinfo{journal}{Nuclear Data Sheets} \bibinfo{volume}{74}
  (\bibinfo{year}{1995}) \bibinfo{pages}{63--164}.
  \DOIprefix\doi{10.1006/ndsh.1994.1047}.
\bibitem[{Huang and Kang(2016)}]{DataSheetsA198}
\bibinfo{author}{X.~Huang}, \bibinfo{author}{M.~Kang},
\newblock \bibinfo{title}{{Nuclear Data Sheets for A = 198}},
\newblock \bibinfo{journal}{Nuclear Data Sheets} \bibinfo{volume}{133}
  (\bibinfo{year}{2016}) \bibinfo{pages}{221--416}.
  \DOIprefix\doi{10.1016/j.nds.2016.02.002}.
\bibitem[{Singh(2007)}]{DataSheetsA64}
\bibinfo{author}{B.~Singh},
\newblock \bibinfo{title}{{Nuclear Data Sheets for A = 64}},
\newblock \bibinfo{journal}{Nuclear Data Sheets} \bibinfo{volume}{108}
  (\bibinfo{year}{2007}) \bibinfo{pages}{197--364}.
  \DOIprefix\doi{10.1016/j.nds.2007.01.003}.
\bibitem[{{Shamsuzzoha Basunia}(2018)}]{DataSheetsA59}
\bibinfo{author}{M.~{Shamsuzzoha Basunia}},
\newblock \bibinfo{title}{{Nuclear Data Sheets for A = 59}},
\newblock \bibinfo{journal}{Nuclear Data Sheets} \bibinfo{volume}{151}
  (\bibinfo{year}{2018}) \bibinfo{pages}{1--333}.
  \DOIprefix\doi{10.1016/j.nds.2018.08.001}.
\bibitem[{Brown \textit{et~al.}(2018)Brown, Chadwick, Capote, Kahler, Trkov,
  Herman, Sonzogni, Danon, Carlson, Dunn, Smith, and Hale}]{ENDF}
\bibinfo{author}{D.~A. Brown}, \bibinfo{author}{M.~B. Chadwick},
  \bibinfo{author}{R.~Capote}, \bibinfo{author}{A.~C. Kahler},
  \bibinfo{author}{A.~Trkov}, \bibinfo{author}{M.~W. Herman},
  \bibinfo{author}{A.~A. Sonzogni}, \bibinfo{author}{Y.~Danon},
  \bibinfo{author}{A.~D. Carlson}, \bibinfo{author}{M.~Dunn},
  \bibinfo{author}{D.~L. Smith}, \bibinfo{author}{G.~M. Hale},
\newblock \bibinfo{title}{{ENDF/B-VIII.0: The 8$^{th}$ Major Release of the
  Nuclear Reaction Data Library with CIELO-project Cross Sections, New Standard
  and Thermal Scattering Data}},
\newblock \bibinfo{journal}{Nuclear Data Sheets} \bibinfo{volume}{148}
  (\bibinfo{year}{2018}) \bibinfo{pages}{1--142}.
  \DOIprefix\doi{10.1016/j.nds.2018.02.001}.
\bibitem[{Trkov \textit{et~al.}(2020)Trkov, Griffin, Simakov, Greenwood,
  Zolotarev, Capote, Aldama, Chechev, Destouches, Kahler, Konno,
  Ko{\v{s}}t{\'{a}}l, Majerle, Malambu, Ohta, Pronyaev, Radulovi{\'{c}}, Sato,
  Schulc, {\v{S}}ime{\v{c}}kov{\'{a}}, Vavtar, Wagemans, White, and
  Yashima}]{IRDFF2}
\bibinfo{author}{A.~Trkov}, \bibinfo{author}{P.~J. Griffin},
  \bibinfo{author}{S.~P. Simakov}, \bibinfo{author}{L.~R. Greenwood},
  \bibinfo{author}{K.~I. Zolotarev}, \bibinfo{author}{R.~Capote},
  \bibinfo{author}{D.~L. Aldama}, \bibinfo{author}{V.~Chechev},
  \bibinfo{author}{C.~Destouches}, \bibinfo{author}{A.~C. Kahler},
  \bibinfo{author}{C.~Konno}, \bibinfo{author}{M.~Ko{\v{s}}t{\'{a}}l},
  \bibinfo{author}{M.~Majerle}, \bibinfo{author}{E.~Malambu},
  \bibinfo{author}{M.~Ohta}, \bibinfo{author}{V.~G. Pronyaev},
  \bibinfo{author}{V.~Radulovi{\'{c}}}, \bibinfo{author}{S.~Sato},
  \bibinfo{author}{M.~Schulc},
  \bibinfo{author}{E.~{\v{S}}ime{\v{c}}kov{\'{a}}},
  \bibinfo{author}{I.~Vavtar}, \bibinfo{author}{J.~Wagemans},
  \bibinfo{author}{M.~White}, \bibinfo{author}{H.~Yashima},
\newblock \bibinfo{title}{{IRDFF-II: A New Neutron Metrology Library}},
\newblock \bibinfo{journal}{Nuclear Data Sheets} \bibinfo{volume}{163}
  (\bibinfo{year}{2020}) \bibinfo{pages}{1--108}.
  \DOIprefix\doi{10.1016/j.nds.2019.12.001}.
  \href{http://arxiv.org/abs/1909.03336}{{\tt arXiv:1909.03336}}.
\bibitem[{Nyarko \textit{et~al.}(2010)Nyarko, Sogbadji, Akaho, and
  Agbemava}]{Nyarko2010}
\bibinfo{author}{B.~J. Nyarko}, \bibinfo{author}{R.~B. Sogbadji},
  \bibinfo{author}{E.~H. Akaho}, \bibinfo{author}{S.~Agbemava},
\newblock \bibinfo{title}{{Cross Section Determination of Short-to-Medium Lived
  Nuclides in a Low Power Research and an Am-Be Neutron Source}},
\newblock in: \bibinfo{booktitle}{Proceedings of the IAEA Technical Meeting in
  collaboration with NEA on Specific Applications of Research Reactors:
  Provision of Nuclear Data (INDC(NDS)-0574)},
  \bibinfo{publisher}{International Atomic Energy Agency},
  \bibinfo{address}{Vienna, Austria}, \bibinfo{year}{2010}, pp.
  \bibinfo{pages}{72--84}. \URLprefix
  \url{http://www-nds.iaea.org/reports-new/indc-reports/indc-nds/indc-nds-0574.pdf}.
\bibitem[{Blaauw \textit{et~al.}(2017)Blaauw, Ridikas, Baytelesov, Salas,
  Chakrova, Eun-Ha, Dahalan, Fortunato, Jacimovic, Kling, Mu{\~{n}}oz, Mohamed,
  P{\'{a}}rk{\'{a}}nyi, Singh, and Duong}]{Blaauw2017}
\bibinfo{author}{M.~Blaauw}, \bibinfo{author}{D.~Ridikas},
  \bibinfo{author}{S.~Baytelesov}, \bibinfo{author}{P.~S. Salas},
  \bibinfo{author}{Y.~Chakrova}, \bibinfo{author}{C.~Eun-Ha},
  \bibinfo{author}{R.~Dahalan}, \bibinfo{author}{A.~H. Fortunato},
  \bibinfo{author}{R.~Jacimovic}, \bibinfo{author}{A.~Kling},
  \bibinfo{author}{L.~Mu{\~{n}}oz}, \bibinfo{author}{N.~M. Mohamed},
  \bibinfo{author}{D.~P{\'{a}}rk{\'{a}}nyi}, \bibinfo{author}{T.~Singh},
  \bibinfo{author}{V.~D. Duong},
\newblock \bibinfo{title}{{Estimation of $^{99}$Mo production rates from
  natural molybdenum in research reactors}},
\newblock \bibinfo{journal}{Journal of Radioanalytical and Nuclear Chemistry}
  \bibinfo{volume}{311} (\bibinfo{year}{2017}) \bibinfo{pages}{409--418}.
  \DOIprefix\doi{10.1007/s10967-016-5036-6}.
\bibitem[{Steinhauser \textit{et~al.}(2012)Steinhauser, Merz, Stadlbauer,
  Kregsamer, Streli, and Villa}]{Steinhauser2012}
\bibinfo{author}{G.~Steinhauser}, \bibinfo{author}{S.~Merz},
  \bibinfo{author}{F.~Stadlbauer}, \bibinfo{author}{P.~Kregsamer},
  \bibinfo{author}{C.~Streli}, \bibinfo{author}{M.~Villa},
\newblock \bibinfo{title}{{Performance and comparison of gold-based neutron
  flux monitors}},
\newblock \bibinfo{journal}{Gold Bulletin} \bibinfo{volume}{45}
  (\bibinfo{year}{2012}) \bibinfo{pages}{17--22}.
  \DOIprefix\doi{10.1007/s13404-011-0039-0}.
\bibitem[{Karadag \textit{et~al.}(2003)Karadag, Y{\"{u}}cel, Tan, and
  {\"{O}}zmen}]{Karadag2003}
\bibinfo{author}{M.~Karadag}, \bibinfo{author}{H.~Y{\"{u}}cel},
  \bibinfo{author}{M.~Tan}, \bibinfo{author}{A.~{\"{O}}zmen},
\newblock \bibinfo{title}{{Measurement of thermal neutron cross-sections and
  resonance integrals for $^{71}$Ga(n,$\gamma$)$^{72}$Ga and
  $^{75}$As(n,$\gamma$)$^{76}$As by using $^{241}$Am-Be isotopic neutron
  source}},
\newblock \bibinfo{journal}{Nuclear Instruments and Methods in Physics Research
  A} \bibinfo{volume}{501} (\bibinfo{year}{2003}) \bibinfo{pages}{524--535}.
  \DOIprefix\doi{10.1016/S0168-9002(03)00408-X}.
\bibitem[{Blaauw(1995)}]{Blaauw1995}
\bibinfo{author}{M.~Blaauw},
\newblock \bibinfo{title}{{The confusing issue of the neutron capture
  cross-section to use in thermal neutron self-shielding computations}},
\newblock \bibinfo{journal}{Nuclear Instruments and Methods in Physics Research
  A} \bibinfo{volume}{356} (\bibinfo{year}{1995}) \bibinfo{pages}{403--407}.
  \DOIprefix\doi{10.1016/0168-9002(94)01316-0}.
\bibitem[{Do \textit{et~al.}(2008)Do, Khue, Thanh, Son, Kim, Lee, Oh, Lee, Cho,
  Ko, and Namkung}]{VanDo2008:W}
\bibinfo{author}{N.~V. Do}, \bibinfo{author}{P.~D. Khue},
  \bibinfo{author}{K.~T. Thanh}, \bibinfo{author}{L.~T. Son},
  \bibinfo{author}{G.~Kim}, \bibinfo{author}{Y.~S. Lee},
  \bibinfo{author}{Y.~Oh}, \bibinfo{author}{H.~Lee}, \bibinfo{author}{M.~Cho},
  \bibinfo{author}{I.~S. Ko}, \bibinfo{author}{W.~Namkung},
\newblock \bibinfo{title}{{Thermal neutron cross-section and resonance integral
  of the $^{186}$W(n,$\gamma$)$^{187}$W reaction}},
\newblock \bibinfo{journal}{Nuclear Instruments and Methods in Physics Research
  B} \bibinfo{volume}{266} (\bibinfo{year}{2008}) \bibinfo{pages}{863--871}.
  \DOIprefix\doi{10.1016/j.nimb.2008.02.021}.
\bibitem[{Do \textit{et~al.}(2015)Do, Khue, Thanh, Hien, Kim, Kim, Shin, Cho,
  and Lee}]{VanDo2015:Sc}
\bibinfo{author}{N.~V. Do}, \bibinfo{author}{P.~D. Khue},
  \bibinfo{author}{K.~T. Thanh}, \bibinfo{author}{N.~T. Hien},
  \bibinfo{author}{G.~Kim}, \bibinfo{author}{K.~Kim}, \bibinfo{author}{S.~G.
  Shin}, \bibinfo{author}{M.~H. Cho}, \bibinfo{author}{M.~Lee},
\newblock \bibinfo{title}{{Thermal neutron capture and resonance integral cross
  sections of $^{45}$Sc}},
\newblock \bibinfo{journal}{Nuclear Instruments and Methods in Physics Research
  B} \bibinfo{volume}{362} (\bibinfo{year}{2015}) \bibinfo{pages}{9--13}.
  \DOIprefix\doi{10.1016/j.nimb.2015.08.075}.
\bibitem[{Mughabghab(2003)}]{Mughabghab2003}
\bibinfo{author}{S.~F. Mughabghab}, \bibinfo{title}{{Thermal Neutron Capture
  Cross Sections, Resonance Integrals, and G-Factors (INDC(NDS)-440)}},
  \bibinfo{type}{Technical Report}, International Nuclear Data Committee, IAEA,
  \bibinfo{address}{Vienna}, \bibinfo{year}{2003}. \URLprefix
  \url{https://www.osti.gov/etdeweb/servlets/purl/20332542}.
\bibitem[{Chilian \textit{et~al.}(2008)Chilian, St-Pierre, and
  Kennedy}]{Chilian2008}
\bibinfo{author}{C.~Chilian}, \bibinfo{author}{J.~St-Pierre},
  \bibinfo{author}{G.~Kennedy},
\newblock \bibinfo{title}{{Complete thermal and epithermal neutron
  self-shielding corrections for NAA using a spreadsheet}},
\newblock \bibinfo{journal}{Journal of Radioanalytical and Nuclear Chemistry}
  \bibinfo{volume}{278} (\bibinfo{year}{2008}) \bibinfo{pages}{745--749}.
  \DOIprefix\doi{10.1007/s10967-008-1604-8}.
\bibitem[{Kumar \textit{et~al.}(2017)Kumar, Sen, Radha, Rao, Acharya, Kumar,
  Venkatasubramani, Reddy, and Joseph}]{Kumar2017}
\bibinfo{author}{G.~V. Kumar}, \bibinfo{author}{S.~Sen},
  \bibinfo{author}{E.~Radha}, \bibinfo{author}{J.~S. Rao},
  \bibinfo{author}{R.~Acharya}, \bibinfo{author}{R.~Kumar},
  \bibinfo{author}{C.~R. Venkatasubramani}, \bibinfo{author}{A.~V. Reddy},
  \bibinfo{author}{M.~Joseph},
\newblock \bibinfo{title}{{Studies on neutron spectrum characterization for the
  Pneumatic Fast Transfer System (PFTS) of KAMINI reactor}},
\newblock \bibinfo{journal}{Applied Radiation and Isotopes}
  \bibinfo{volume}{124} (\bibinfo{year}{2017}) \bibinfo{pages}{49--55}.
  \DOIprefix\doi{10.1016/j.apradiso.2017.03.009}.
\bibitem[{Dung and Sasajima(2003)}]{Dung2003}
\bibinfo{author}{H.~M. Dung}, \bibinfo{author}{F.~Sasajima},
\newblock \bibinfo{title}{{Determination of $\alpha$ and f for k$_0$-NAA in
  irradiation sites with high thermalized neutrons}},
\newblock \bibinfo{journal}{Journal of Radioanalytical and Nuclear Chemistry}
  \bibinfo{volume}{257} (\bibinfo{year}{2003}) \bibinfo{pages}{509--512}.
  \DOIprefix\doi{10.1023/A:1025472011260}.
\bibitem[{{De Corte} \textit{et~al.}(1989){De Corte}, Simonits, and
  Wispelaere}]{DeCorte1989}
\bibinfo{author}{F.~{De Corte}}, \bibinfo{author}{A.~Simonits},
  \bibinfo{author}{A.~D. Wispelaere},
\newblock \bibinfo{title}{{Comparative Study of Measured and Critically
  Evaluated Resonance Integral To Thermal Cross Section Ratios, III}},
\newblock \bibinfo{journal}{Journal of Radioanalytical and Chemistry}
  \bibinfo{volume}{133} (\bibinfo{year}{1989}) \bibinfo{pages}{131--151}.
  \DOIprefix\doi{10.1007/BF02039971}.
\bibitem[{Y{\"{u}}cel and Karadag(2004)}]{Yucel2004}
\bibinfo{author}{H.~Y{\"{u}}cel}, \bibinfo{author}{M.~Karadag},
\newblock \bibinfo{title}{{Experimental determination of the $\alpha$-shape
  factor in the 1/E$^{1+\alpha}$ epithermal-isotopic neutron source-spectrum by
  dual monitor method}},
\newblock \bibinfo{journal}{Annals of Nuclear Energy} \bibinfo{volume}{31}
  (\bibinfo{year}{2004}) \bibinfo{pages}{681--695}.
  \DOIprefix\doi{10.1016/j.anucene.2003.10.002}.
\bibitem[{{De Corte} \textit{et~al.}(1979){De Corte}, Moens, Simonits, {De
  Wispelaere}, and Hoste}]{DeCorte1979}
\bibinfo{author}{F.~{De Corte}}, \bibinfo{author}{L.~Moens},
  \bibinfo{author}{A.~Simonits}, \bibinfo{author}{A.~{De Wispelaere}},
  \bibinfo{author}{J.~Hoste},
\newblock \bibinfo{title}{{Instantaneous $\alpha$-determination without
  Cd-cover in the 1/E$^{1+\alpha}$ epithermal neutron spectrum}},
\newblock \bibinfo{journal}{Journal of Radioanalytical Chemistry}
  \bibinfo{volume}{52} (\bibinfo{year}{1979}) \bibinfo{pages}{295--304}.
  \DOIprefix\doi{10.1007/BF02521280}.
\bibitem[{Eastwood and Werner(1962)}]{Eastwood1962}
\bibinfo{author}{T.~A. Eastwood}, \bibinfo{author}{R.~D. Werner},
\newblock \bibinfo{title}{{Resonance and Thermal Neutron Self-Shielding in
  Cobalt Foils and Wires}},
\newblock \bibinfo{journal}{Nuclear Science and Engineering}
  \bibinfo{volume}{13} (\bibinfo{year}{1962}) \bibinfo{pages}{385--390}.
  \DOIprefix\doi{10.13182/nse62-a26181}.
\bibitem[{Jovanovic \textit{et~al.}(1985)Jovanovic, Corte, Simonits, Moens,
  Vukotic, Zejnilovic, and Hoste}]{Jovanovic1985}
\bibinfo{author}{S.~Jovanovic}, \bibinfo{author}{F.~D. Corte},
  \bibinfo{author}{A.~Simonits}, \bibinfo{author}{L.~Moens},
  \bibinfo{author}{P.~Vukotic}, \bibinfo{author}{R.~Zejnilovic},
  \bibinfo{author}{J.~Hoste},
\newblock \bibinfo{title}{{Epithermal Neutron Flux Distribution and Its Impact
  on (n,$\gamma$) Activation Analysis Result}},
\newblock in: \bibinfo{booktitle}{First Balkan conference on activation
  analysis}, \bibinfo{address}{Bulgaria}, \bibinfo{year}{1985}, p.
  \bibinfo{pages}{196}. \URLprefix
  \url{https://inis.iaea.org/search/search.aspx?orig_q=RN:19030914}.
\bibitem[{Martinho \textit{et~al.}(2003)Martinho, Gon{\c{c}}alves, and
  Salgado}]{Martinho2003}
\bibinfo{author}{E.~Martinho}, \bibinfo{author}{I.~F. Gon{\c{c}}alves},
  \bibinfo{author}{J.~Salgado},
\newblock \bibinfo{title}{{Universal curve of epithermal neutron resonance
  self-shielding factors in foils, wires and spheres}},
\newblock \bibinfo{journal}{Applied Radiation and Isotopes}
  \bibinfo{volume}{58} (\bibinfo{year}{2003}) \bibinfo{pages}{371--375}.
  \DOIprefix\doi{10.1016/S0969-8043(02)00313-5}.
\bibitem[{Qaim \textit{et~al.}(1988)Qaim, Mushtaq, and
  Uhl}]{Qaim1988:73SeIsomer}
\bibinfo{author}{S.~M. Qaim}, \bibinfo{author}{A.~Mushtaq},
  \bibinfo{author}{M.~Uhl},
\newblock \bibinfo{title}{{Isomeric cross-section ratio for the formation of
  $^{73m,g}$Se in various nuclear processes}},
\newblock \bibinfo{journal}{Physical Review C} \bibinfo{volume}{38}
  (\bibinfo{year}{1988}) \bibinfo{pages}{645--650}.
  \DOIprefix\doi{10.1103/PhysRevC.38.645}.
\bibitem[{Marsh(1836)}]{MarshTest}
\bibinfo{author}{J.~Marsh},
\newblock \bibinfo{title}{{Account of a method of separating small quantities
  of arsenic from substances with which it may be mixed}},
\newblock \bibinfo{journal}{Edinburgh New Philosophical Journal}
  \bibinfo{volume}{21} (\bibinfo{year}{1836}) \bibinfo{pages}{229--236}.
\bibitem[{Esposito \textit{et~al.}(2019)Esposito, Bettoni, Boschi, Calderolla,
  Cisternino, Fiorentini, Keppel, Martini, Maggiore, Mou, Pasquali, Pranovi,
  Pupillo, Alvarez, Sarchiapone, Sciacca, Skliarova, Favaron, Lombardi,
  Antonini, and Duatti}]{Esposito2019}
\bibinfo{author}{J.~Esposito}, \bibinfo{author}{D.~Bettoni},
  \bibinfo{author}{A.~Boschi}, \bibinfo{author}{M.~Calderolla},
  \bibinfo{author}{S.~Cisternino}, \bibinfo{author}{G.~Fiorentini},
  \bibinfo{author}{G.~Keppel}, \bibinfo{author}{P.~Martini},
  \bibinfo{author}{M.~Maggiore}, \bibinfo{author}{L.~Mou},
  \bibinfo{author}{M.~Pasquali}, \bibinfo{author}{L.~Pranovi},
  \bibinfo{author}{G.~Pupillo}, \bibinfo{author}{C.~R. Alvarez},
  \bibinfo{author}{L.~Sarchiapone}, \bibinfo{author}{G.~Sciacca},
  \bibinfo{author}{H.~Skliarova}, \bibinfo{author}{P.~Favaron},
  \bibinfo{author}{A.~Lombardi}, \bibinfo{author}{P.~Antonini},
  \bibinfo{author}{A.~Duatti},
\newblock \bibinfo{title}{{LARAMED: A Laboratory for Radioisotopes of Medical
  Interest}},
\newblock \bibinfo{journal}{Molecules} \bibinfo{volume}{24}
  (\bibinfo{year}{2019}) \bibinfo{pages}{20}.
  \DOIprefix\doi{10.3390/molecules24010020}.
\bibitem[{Sukhoruchkin and Soroko(2015)}]{Sukhoruchkin2015}
\bibinfo{author}{S.~I. Sukhoruchkin}, \bibinfo{author}{Z.~N. Soroko},
  \bibinfo{title}{{Neutron Resonance Parameters for As-75 (Arsenic)}},
  \bibinfo{publisher}{Springer Materials}, \bibinfo{year}{2015}.
  \DOIprefix\doi{10.1007/978-3-662-45603-3}.
\bibitem[{Mughabghab and Garber(1973)}]{Mughabghab1973}
\bibinfo{author}{S.~Mughabghab}, \bibinfo{author}{D.~Garber},
  \bibinfo{title}{{Neutron Cross Sections. Volume 1, Resonance Parameters
  (BNL-325; EANDC(US)-183/L; INDC(USA)-58/L)}}, \bibinfo{type}{Technical
  Report}, National Neutron Cross Section Center, Brookhaven National
  Laboratory, \bibinfo{address}{Upton, New York}, \bibinfo{year}{1973}.
  \DOIprefix\doi{10.2172/4335199}.

\end{thebibliography}

\end{document}